\documentclass[11pt]{article}

\title{Topological materials with extensive flat-band surface states}
\author{Protyush Nandi$^1$ and Subinay Dasgupta$^2$\\ 
$^1$Department of Physics, University of Calcutta,92 Acharya Prafulla Chandra Road,\\ Kolkata 700009, India\\
$^2$Department of Physics, Harish-Chandra Research Institute, Prayagraj 211019, India}
\date{}

\usepackage[top=.5 in,bottom=1 in, left=1 in, right=1 in]{geometry}
\usepackage[backend=biber,style=nature,]{biblatex}
\usepackage{color}
\usepackage{verbatim}
\usepackage{amssymb}
\usepackage{amsmath}
\usepackage{epsfig}
\usepackage{graphics}
\usepackage{graphicx}
\usepackage{dcolumn}
\usepackage{bm}
\usepackage{float}
\usepackage{subcaption}

\def\be{\begin{equation}}
\def\ee{\end{equation}}
\def\bea{\begin{eqnarray}}
\def\eea{\end{eqnarray}}
\def\bfg{\begin{figure}[H]}
\def\efg{\end{figure}}

\begin{document}

\maketitle

\begin{abstract}
Materials that have zero-energy flat band states on the surface may show surface superconductivity. Here we report a theoretical observation that a Hamiltonian describing a thin slab of topological nodal line semimetal, has zero energy eigenstate spanning the entire surface of the Brillouin zone under certain conditions, namely (i) the {hopping amplitude} of fermions in the direction of thickness is more than that in other directions (ii) the {onsite energy} should be less than some limiting value determined by the hopping probability. Our claim is substantiated by analytic and numerical approach. We also report new phase transitions in a region of parameter space and indicate that the Hamiltonian can also be realised by stacked layers {described by a suitable Hamiltonian.}
\end{abstract}

Topological nodal line semimetal (TNLSM) is a type of topological material characterised by the presence of nodal points (where the conduction and valence bands touch) along a closed loop in the momentum space. These materials have a unique importance due to their distinctive properties, like flat drumhead states \cite{Weng, Yu, Kim}, graphene-like transport properties in 3D \cite{Burkov2011} and plasmon mode behaviour \cite{Yan, Rhim-Kim}. Bulk superconductivity has been observed in several such materials, such as, CaSb$_2$ \cite{Duan2022}, SnTaS$_2$ \cite{Singh2022}, NaAlSi \cite{Hirai2022}, PbTaSe$_2$ \cite{Bian} .  Surface superconductivity in a topological flat band system has been predicted long back  \cite{Kopnin,Kopnin2011supercur,Heikkila2011flat,Volovik2013media,Esquinazi2014,Tang2014}, and  has been observed for graphene multilayers \cite{Zhang2022} and the TNLSM system Pd-doped CaAgP \cite{Yano2023}.  Surface superconductivity has also been observed for 3D Dirac material BaMg$_2$Bi$_2$ \cite{Liu2022} and Weyl semimetal PtBi$_2$ \cite{Kuibarov2024}. Furthermore, it has been shown by {\em ab initio} calculations \cite{Electride} that the topological materials Y$_2$C and Sc$_2$C should have zero energy states covering extended regions of Brilloun zone surface.

{In this article, we report a reasonable Hamiltonian which shows flat band {(i.e. dispersionless)} surface states on the top and bottom surfaces of a sample with thin slab geometry, encompassing the entire X-Y plane in momentum space.} For an earlier version (without the parameter $t$, and the discussion on extensive flat band states) see \cite{nandi2023node}. Thus, a material prepared to be represented by this Hamiltonian will have the prospect of having prominent surface superconductivity {with a high critical temperature. This model shows a rich phase diagram consisting of a number of topological  phases.} The Hamiltonian under question in the 3D wave-vector space (with periodic boundary condition) is constituted of $2\times 2$ {{\em free spinless Fermions}} in $\vec{k}$ space:
  \bea {\mathcal H} = \sum_{\vec{k}} c_{\vec{k}}^{\dagger} {\mathcal{H}}_{\vec{k}} c_{\vec{k}} \;\;\;\;\;\; {\mathcal H}_{\vec{k}} = A_{\vec{k}} \sigma_z + B_{\vec{k}} \sigma_x \label{def_Hk}\eea
 with 
 \bea \label{akbk}
 A_{\vec{k}}  = J - t\cos k_x - t\cos k_y - \cos k_z \\B_{\vec{k}}  =  t\sin k_x + t\sin k_y + \sin k_z \nonumber  
 \eea
 \\
 where, $c_{\vec{k}}^{\dagger}=(c_{\vec{k}A}^{\dagger} \;\;\; c_{\vec{k}B}^{\dagger})$, $c_{\vec{k}}=(c_{\vec{k}A} \;\;\; c_{\vec{k}B})$ denote the creation and annihilation operators for two sublattices $A$ and $B$, $J$ and $t$  are parameters and $\vec{k}$ runs over the first Brillouin zone (BZ) of a cubic lattice $-\pi < k_x, k_y, k_z \le \pi$. (The motivations for this Hamiltonian will be discussed below.) Note that this Hamiltonian is anisotropic in the Z direction (unless $t=1$) 
 \begin{figure*}
 \centering
    \includegraphics[scale=0.35]{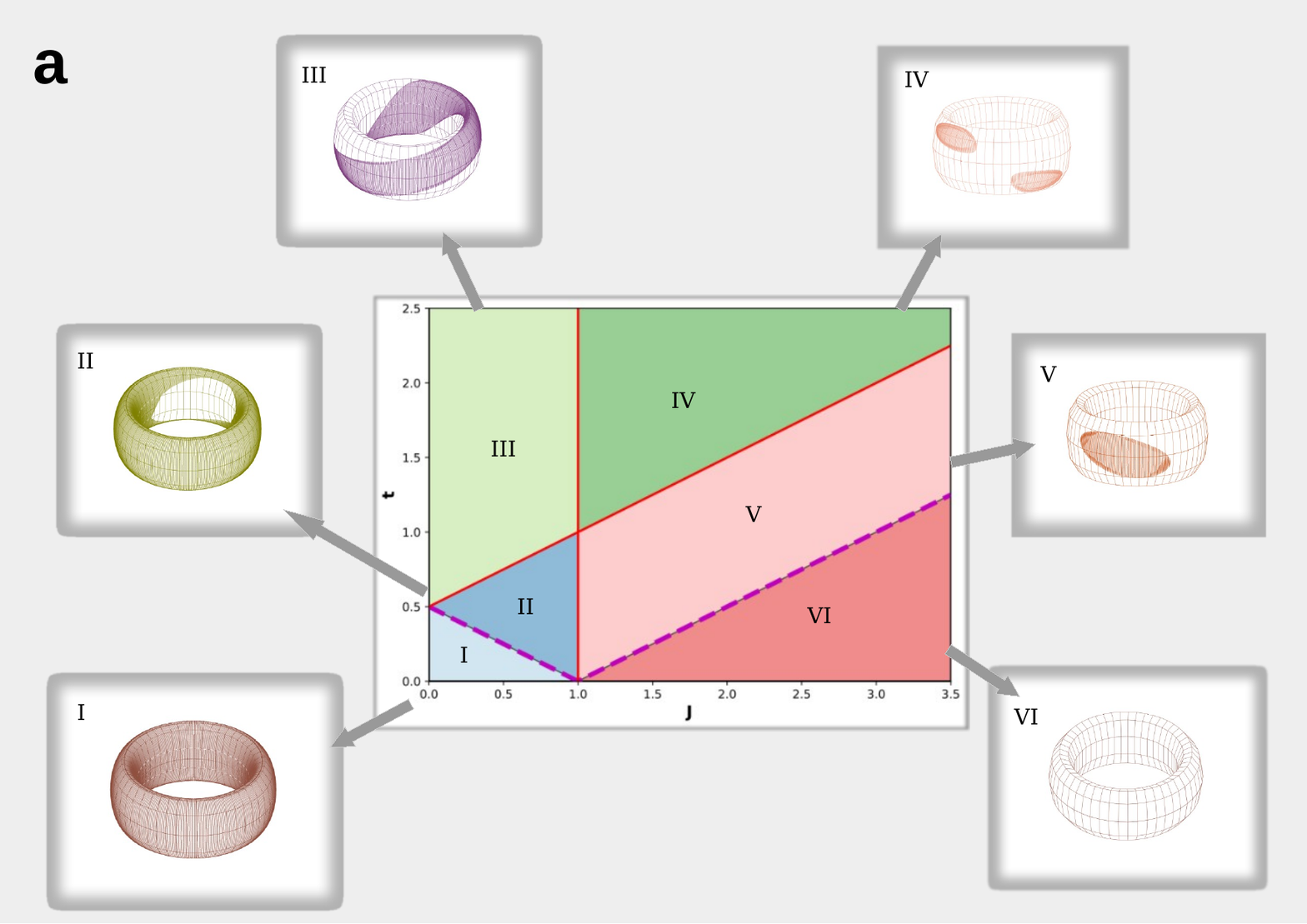}
    \includegraphics[scale=0.35]{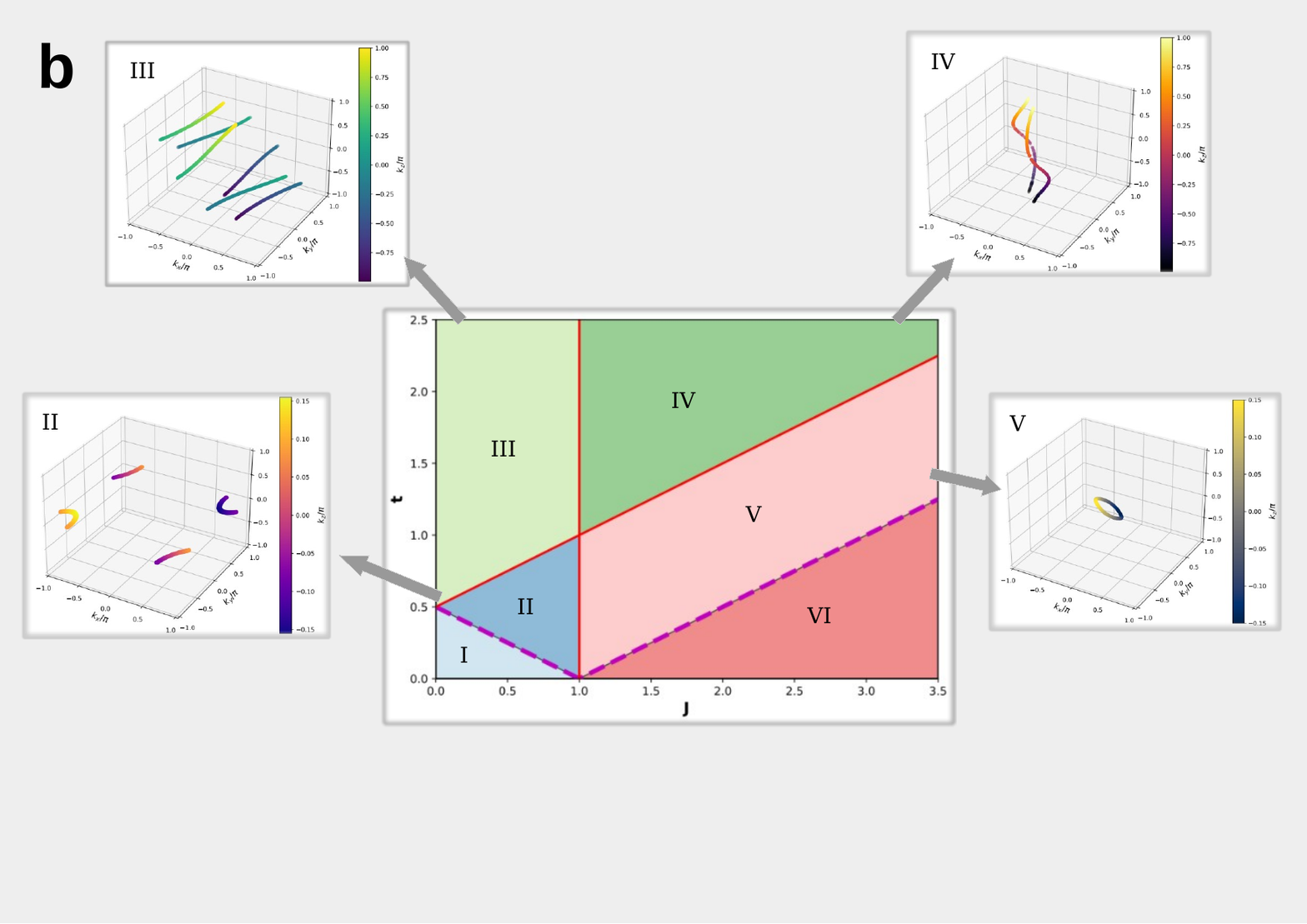}
    \caption{\small{The phase diagram (at zero temperature)  in $t-J$ plane. There are six phases, irrespective of whether we have open or closed boundary in the Z direction. Fig. 1a is for periodic boundary in X and Y direction and open boundary in Z direction. We show here the regions on the toroidal $k_x-k_y$ surface at which zero-energy states, localized at surfaces $n_z=1$ and $n_z=L_z$, appear. Fig. 1b is for fully periodic boundaries. We show here the nodal lines (where the energy is zero) in the $k_x-k_y-k_z$ Brillouin zone. In both the figures, the nature of non-analyticity at the phase boundary is indicated by colour, dashed magenta corresponds to divergence of $\partial^3 E_0/\partial t^3$ and red, to a discontinuous change in the same quantity.
    (For computing the location of nodal lines we used $N=400^3$. For computing the surface states, we used $L_z=200$)\\ 
    Phase I: Topological insulator with zero-energy surface state  ranging all over the $k_x-k_y$ plane.  \\
    Phase II: Nodal line semimetal phase, with one nodal line, which touches $k_x-k_z$ and $k_y-k_z$ boundary surfaces. The zero-energy states form a multiply connected region, but the non-zero energy states form a simply connected region (a patch). \\
    Phase III: Another nodal line phase, but with two nodal lines, which touch all the surfaces. Both the zero-energy  and non-zero-energy states form multiply connected regions. \\
    Phase IV: Two nodal lines in this phase touch only the $k_x-k_y$ boundary surfaces, but not the $k_x-k_z$ and $k_y-k_z$ surfaces. The non-zero-energy states form a multiply connected region, but the zero-energy states form two patches.\\
    Phase V: This phase contains only one nodal loop, which does not touch any boundary surface of the BZ. The non-zero-energy states form a multiply connected region, but the zero-energy states form one patch.\\
   Phase VI: This is a normal insulator phase without any node and zero-energy state on surface.\\
    {The equations of critical lines separating a) II and V, III and IV is $J=1$, b) II and III, V and IV is $J-2t=-1$ c) I and II is $J+2t=1$ d) V and VI is $J-2t=1$}} }      
    \label{PhaseDiag}
\end{figure*}
 The corresponding Hamiltonian in real space (with periodic boundary condition) is (Eq. \ref{H_real}),
\bea 
\mathcal{H}=\sum_{n_x=1}^{L_x}  \sum_{n_y=1}^{L_y} \sum_{n_z=1}^{L_z} \left[ c_{n_x n_y n_z }^{\dagger} J \sigma_z c_{n_x n_y n_z } + tc_{n_x n_y n_z }^{\dagger} \sigma^{\pm} c_{n_x n_y\pm 1\, n_z } + tc_{n_x n_y n_z }^{\dagger} \sigma^{\pm} c_{n_x\pm 1\, n_y n_z } \right. \nonumber \\ 
\left. + c_{n_x n_y n_z }^{\dagger} \sigma^{\pm} c_{n_x n_y n_z\pm 1 } \right]
\label{H_real}   
\eea
where $\sigma^{\pm} = -(\sigma_z \pm i\sigma_x)/2$. This Hamiltonian represents a system with on-site energy $J$ and hops to and from the nearest neighbouring sites along X, Y, Z axes with amplitudes $t$, $t$, 1 respectively. To adapt the forms of Eqs (\ref{def_Hk},\ref{H_real}) to 
the slab geometry, we perform an inverse Fourier transform of the BZ Hamiltonian Eq. (\ref{def_Hk}) in the Z direction and set open boundary condition in that direction. An alternative is to perform Fourier transformation of the direct space Hamiltonian Eq. (\ref{H_real}) in X and Y directions and impose open boundary condition along Z direction.  This gives 
 \bea 
 \mathcal{H}=\sum_{k_x, k_y} \sum_{n_z=1}^{L_z} \left[ c_{k_x k_y n_z }^{\dagger} \left( \Delta_1 \sigma_z +  \Delta_2 \sigma_x \right) c_{k_x k_y n_z }
+ c_{k_x k_y n_z }^{\dagger} \sigma^+ c_{k_x k_y n_z+1 }
+ c_{k_x k_y n_z }^{\dagger} \sigma^- c_{k_x k_y n_z-1 } \right] \label{H_slab} 
\eea
where $\Delta_1 = J- t\cos(k_x) - t\cos(k_y)$, and $\Delta_2 = t \sin(k_x)+t \sin(k_y)$. (For each $n_z$, the $k_x-k_y$ surface is then a toroid.) A similar Hamiltonian has been proposed for layered materials by Burkov, Hook and Balents \cite{Burkov2011}.  The crucial result of this paper is that {in Phase I} the zero energy eigenstates of this Hamiltonian  (Eq (\ref{H_slab})),{which are localized on the surfaces $n_z=1$ and $n_z=L_z$}, cover the entire $k_x-k_y$ face of the BZ for the parameter region $0<J<1$, $0<t<(1-J)/2$ and form a flat band. Hence, a material described by our Hamiltonian is likely to show high-temperature surface superconductivity, as predicted elsewhere \cite{Kopnin}. Furthermore, the existence of flat bands all-over the surface, is not destroyed if the hopping amplitude in Z direction fluctuates around 1 (and is uniformly distributed between 0.9 and 1.1). {We want to emphasise the fact that while flat band surface states can arise due to anisotropic hopping irrespective of the bulk topology, the flat band surface states in our model are topologically protected and robust (Fig. \ref{disorder_dipsersion}).}\\

{\bf Motivations for the Hamiltonian:} We have mentioned above that our  Hamiltonian (\ref{H_real}) has been used to describe layered materials. For the special case of $t=1$, there are several other motivations: \\
(i) On the plane $k_x = -k_y$, for small $\vec{k}$, our Hamiltonian of Eq. (\ref{def_Hk}) reduces to the usual Hamiltonian for TNLSM upto terms quadratic in $\vec{k}$ \cite{Shen}, \cite{Armitage}, \cite{Yang2022}.
\be  {\mathcal H}_{\vec{k}}^0 = vk_z \sigma_x + (k^2 - k_0^2) \sigma_z \label{usual_Hk}  \ee
(Here $v$ and $k_0$ are the parameters of the model and insulator to semimetal transition occurs as the parameter $k_0$ is varied.) \\
 (ii) The usual Hamiltonian for Weyl semimetal \cite{Shen, Armitage, Paper1, Paper2} 
 \bea {\mathcal H}_{\vec{k}}^W = (J - \cos k_x - \cos k_y - \cos k_z) \sigma_z \\ +\sin k_x  \sigma_x + \sin k_y  \sigma_y \nonumber \eea
 has a form similar to Eq. (\ref{def_Hk}).\\
(iii) The matrix representation for the Hamiltonian (\ref{H_slab}) has a tridiagonal form 
\be \left( \begin {array}{cccccc}
S & \sigma^- & 0 & 0 & \cdots & 0  \\
\sigma^+ & S & \sigma^- & 0 & \cdots & 0 \\
0 & \sigma^+ & S & \sigma^- & \cdots & 0 \\
\cdots &  &  &  & & \\
0 & 0 & 0 & \cdots & \sigma^+  & S
\end{array}      \right) \label{matrix} \ee
where $S = \Delta_1 \sigma_z + \Delta_2 \sigma_x $. 
Since $S$ corresponds to {a toy model of Dirac semimetal}, our Hamiltonian can also be realised by stacking layers of {Dirac semimetals} with intralayer hopping probability $t$ and interlayer hopping probability unity.
\\ \\
{\bf Symmetries of the Hamiltonian:} The time reversal operator is ${\mathcal T} = \sigma_z {\mathcal K}$
 (where ${\mathcal K}$ is complex conjugation operator \cite{Chiu,Fang_2016}) and we observe that,
 {\bea 
 {\mathcal T} (A_{\vec{k}} \sigma_z + B_{\vec{k}} \sigma_x) 
 {\mathcal T}^{-1} & = & A_{-\vec{k}}\sigma_z + B_{-\vec{k}} \sigma_x
 \nonumber \\
 {\mathcal T} \left(\sum_{\vec{k}} {\mathcal H}_{\vec{k}}\right)  
  {\mathcal T}^{-1} & = & \sum_{\vec{k}} {\mathcal H}_{\vec{k}} \eea}
 Since the Brillouin zone is symmetric about $\vec{k}=0$, corresponding to every $\vec{k}$ in the sum of Eq. (\ref{def_Hk}) there will be a $-\vec{k}$. Hence our Hamiltonian is time-reversal symmetric.
The Hamiltonian is also invariant under inversion symmetry $\mathcal{P}=\sigma_z$  since,
 ${\sigma_z} \mathcal{H}(-\vec{k}) {\sigma_z} = \mathcal{H}(\vec{k})$.
There is sub-lattice symmetry also, since $\sigma_y \mathcal{H}(\vec{k}) \sigma_y=-\mathcal{H}(\vec{k})$.\\

{The $\mathcal{H}_{\vec{k}}$ of the model under study is a real Hamiltonian, which is a consequence of the presence of both time reversal and inversion symmetry or $\mathcal{P}*\mathcal{T}$ symmetry. In a $\mathcal{P}*\mathcal{T}$ symmetric system, Berry curvature vanishes except at degeneracy points, from which we can conclude that the nodal lines in this model has $\pi$ Berry phase \cite{Fang_2016}. }

\begin{figure}
\centering
{\includegraphics[scale=0.45]{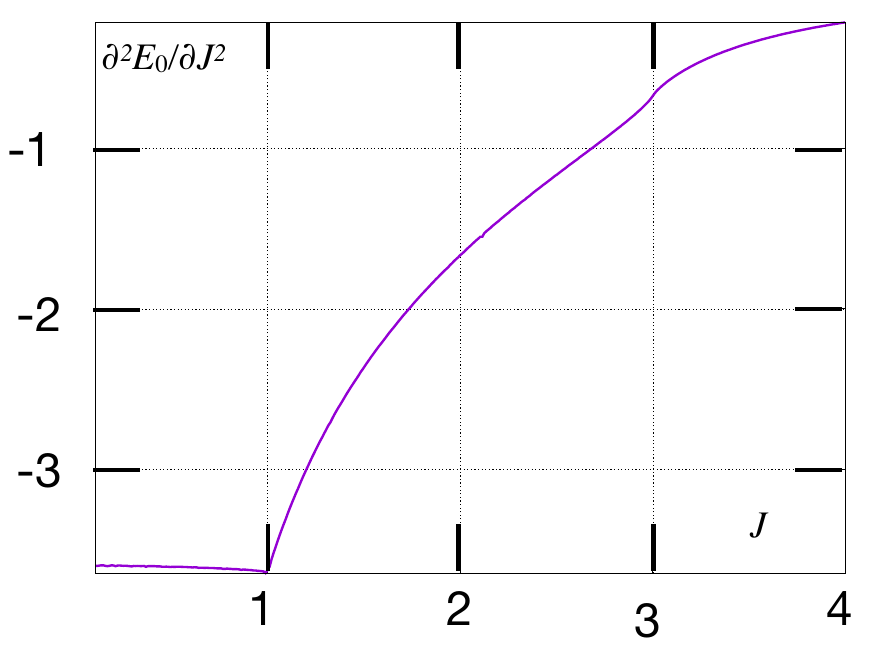}}
{\includegraphics[scale=0.45]{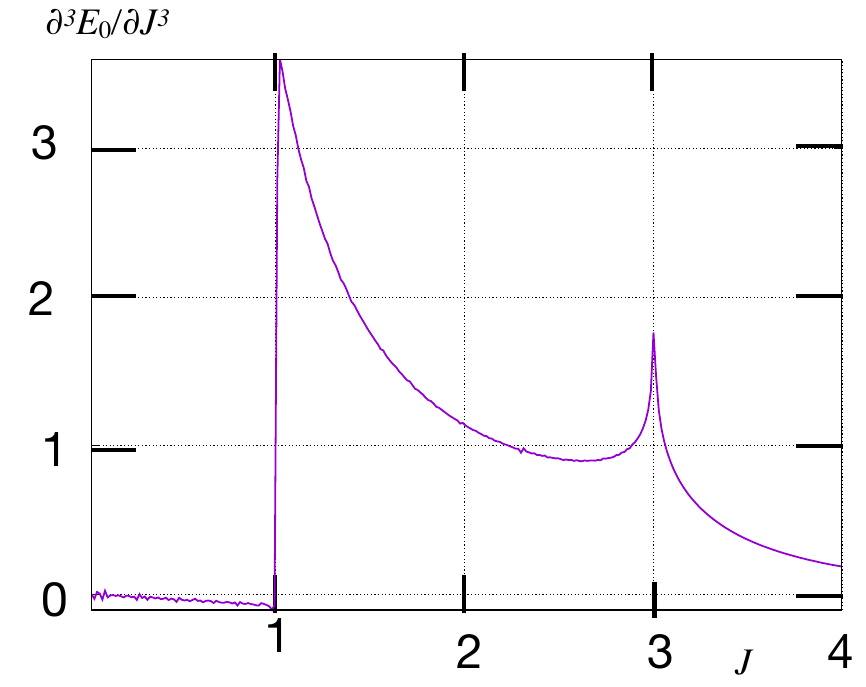}}
\caption{Derivatives of ground state energy  calculated from Eq. (\ref{E_0(J,t)}) for $t=1$ and $N=300^3$.}
\label{Derivatives}
\end{figure}

{\bf  Nodal Lines and Phase Diagram:} Nodal lines consist of the points in the 3D Brillouin zone at which energy is zero, that is, $A_{\vec{k}}=0$ and $B_{\vec{k}}=0$, with $A_{\vec{k}}$, $B_{\vec{k}}$ given by Eq.(\ref{akbk}). For a given $k_z$, one obtains,
\bea 2t\cos\frac{k_x + k_y}{2} \cos\frac{k_x - k_y}{2} & = & J - \cos k_z \nonumber \\ 
2t\sin\frac{k_x + k_y}{2} \cos\frac{k_x - k_y}{2} & = & -\sin k_z.  \eea
and one can solve out $k_x$, $k_y$ from here.  
The ground state energy (per site)  of the Hamiltonian in Eq. (\ref{akbk}) is
\be E_0(J,t) = - \frac{1}{N} \sum_{\vec{k}} \sqrt{A_{\vec{k}}^2 + B_{\vec{k}}^2}  \label{E_0(J,t)}\ee
where $N$ is the total number of sites. This quantity shows non-analytic behaviour as a function of $J$ and $t$ along some lines in the $J-t$ plane. These lines mark the boundaries of different phases (Fig. \ref{PhaseDiag}). The singularity for $t=1$ is presented in Fig. \ref{Derivatives}. There are six such phases, each characterised by specific topologies of nodal lines. The phase boundaries are also characterised by different types of non-analyticity in the ground state energy. Similar transitions between TNLSM phases have been observed earlier \cite{Jiang2018}. \\ 
{Topological phase transitions between two nodal line phases are usually of two types, characterised by (a) a change in linking structure of the nodal lines \cite{Ezawa2017} and (b) a change in the number of nodal lines in the bulk \cite{Jiang2018}. In our phase diagram, the transition between the phases II and V do not fall in these categories, since these two phases contain one nodal line, the same linking topology, and the same symmetry. The transition is only accompanied by a change in topological character of the nodal lines and of the region occupied by the surface states. Similarly, the transition between the phases III and IV are also unique, since both the phases have two nodal lines, the same linking topology, and the same symmetry and differ as regards the topology of the nodal lines and of the region occupied by the surface states.\\}

\begin{figure}
    \centering
    \includegraphics[scale=0.45]{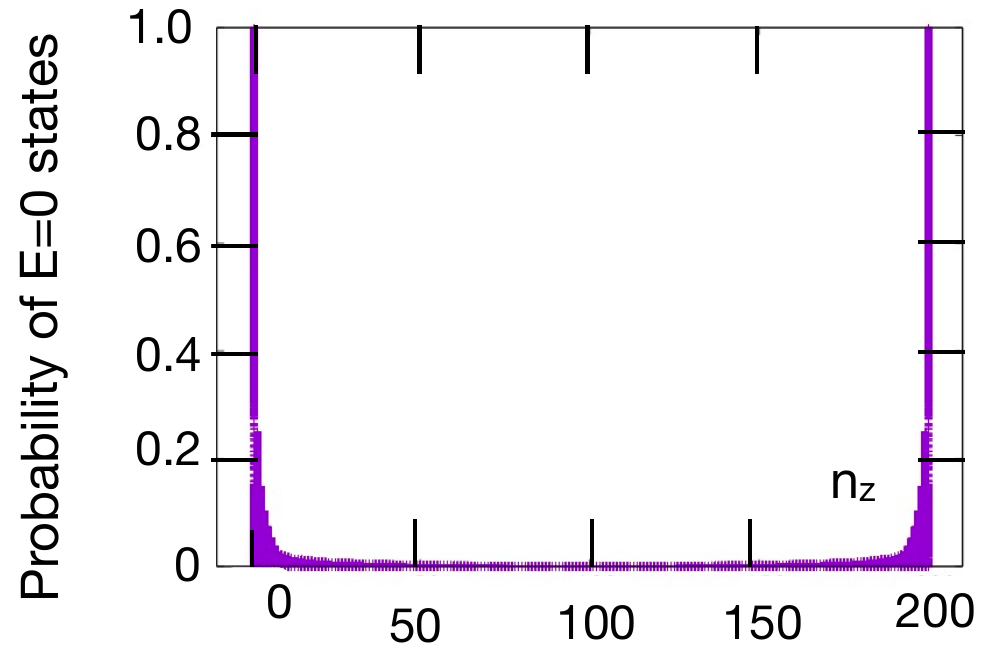}
    \includegraphics[scale=0.45]{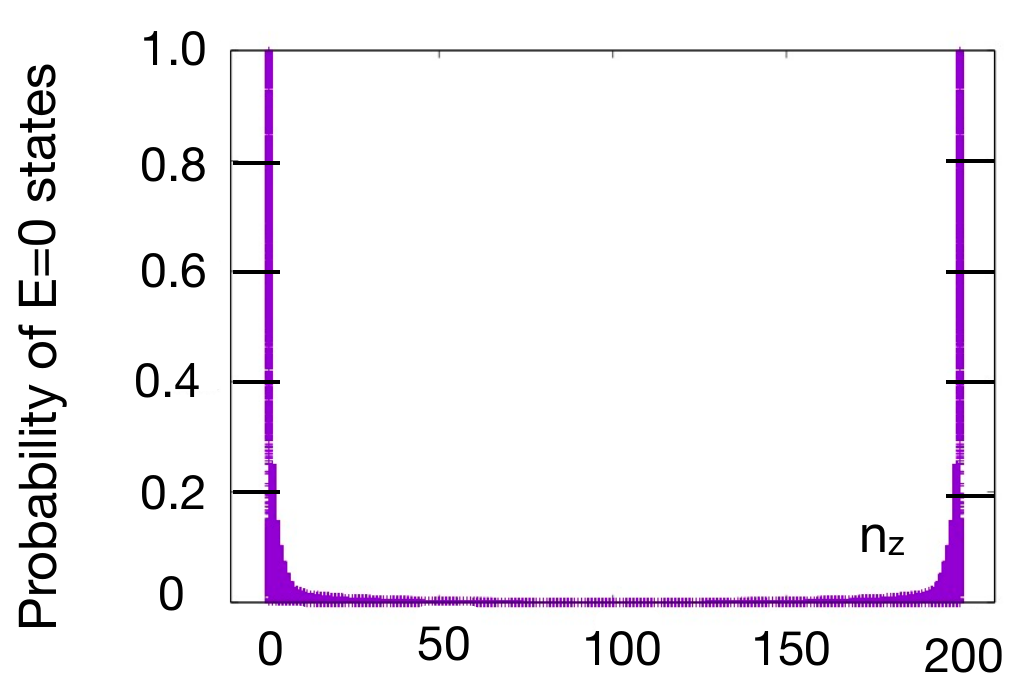}
    \caption{The probability distribution of all the $E=0$ states vs $n_z$ in phase III(left) and phase V(right).}
    \label{edge_prob}
\end{figure}

{\bf Surface States:} {Let the eigenstates of the slab-Hamiltonian Eq. (\ref{H_slab}, \ref{matrix}) be $\psi_{n}(k_x, k_y)$ with eigenvalues $E_{n}$. We identify, numerically, those points $(k_x, k_y)$ for which there exists at least one ${n}$ with $E_{n}=0$ and with large values of  $|\psi_n(k_x, k_y)|^2$ at $n_z=1$ and $n_z=L_z$ (Fig. \ref{edge_prob}). Fig. \ref{PhaseDiag} describes the different types of surface states observed numerically in the six phases.\\}

From an analytic viewpoint, one finds that the Hamiltonian in Eq. (\ref{matrix}) shows surface states when the condition 
\be \Delta_1^2 + \Delta_2^2 < 1 \label{SSH_condn} \ee
is satisfied (see Appendix section for details). \\

\begin{figure*}
\centering
\includegraphics[scale=0.35]{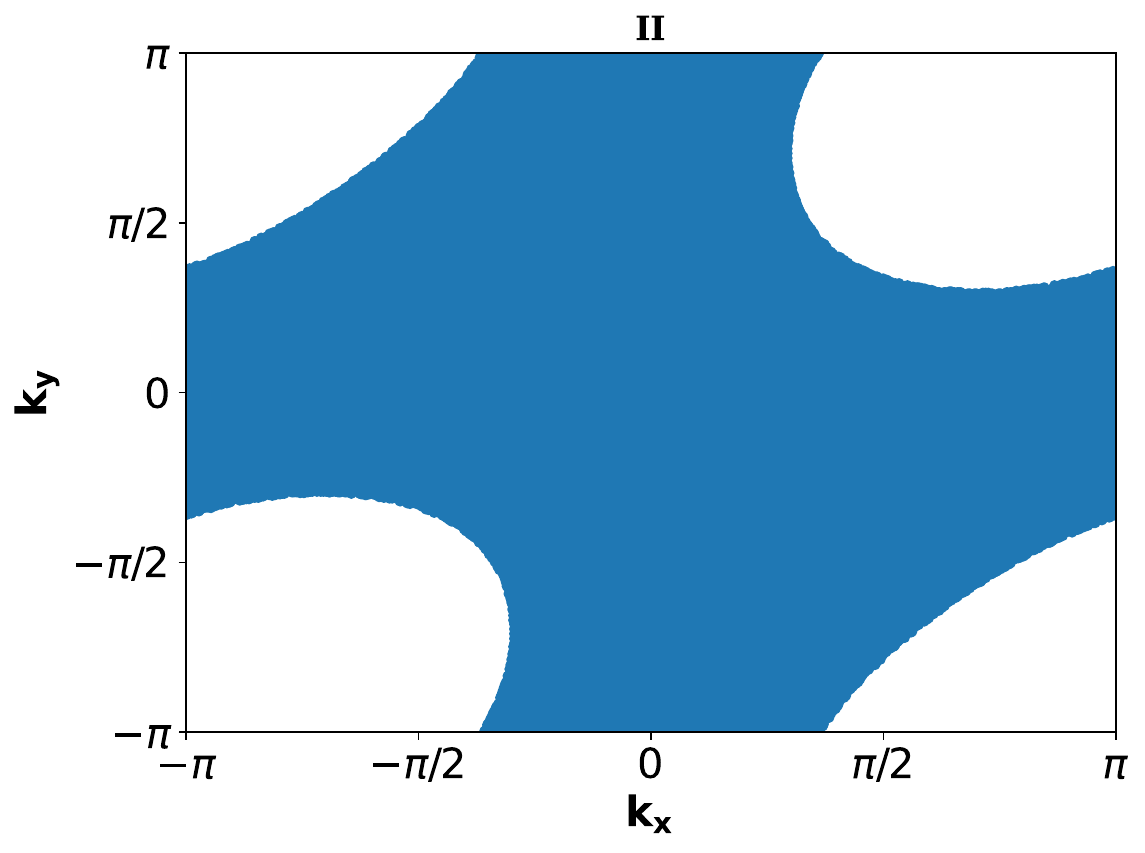}
 \includegraphics[scale=0.35]{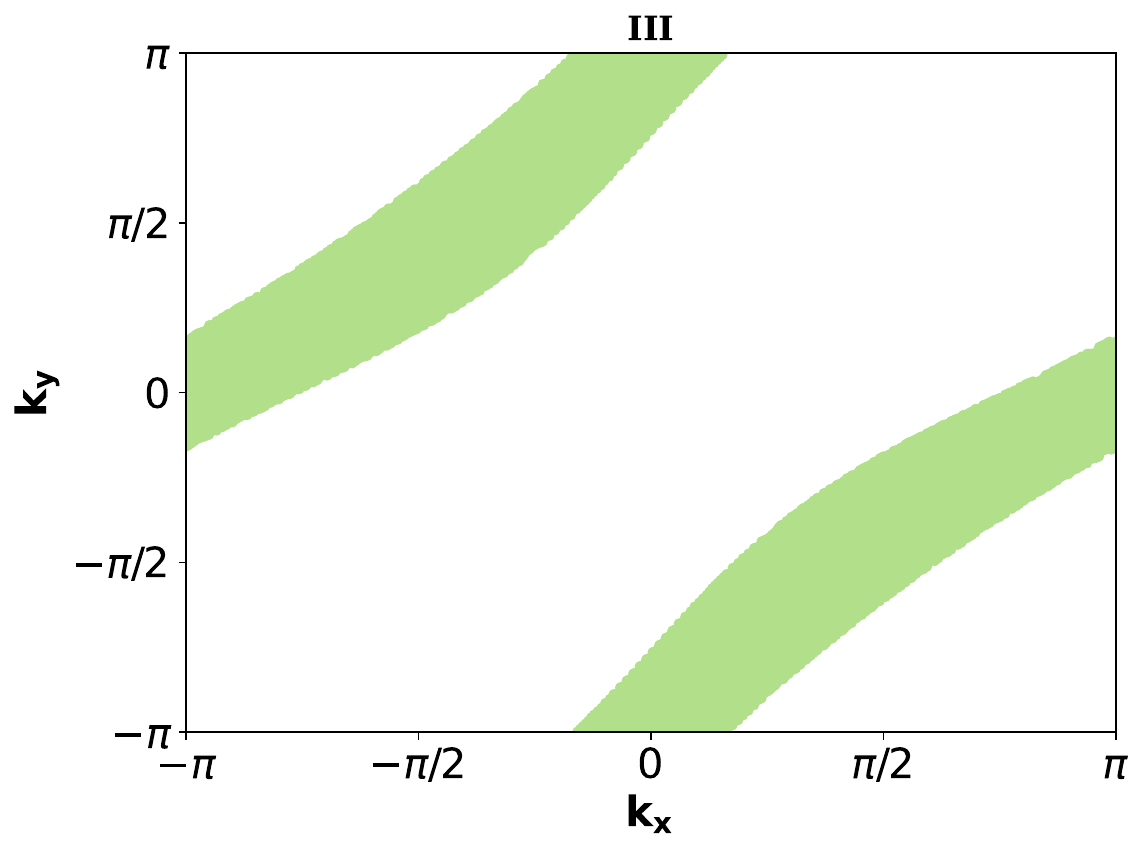}
 \includegraphics[scale=0.35]{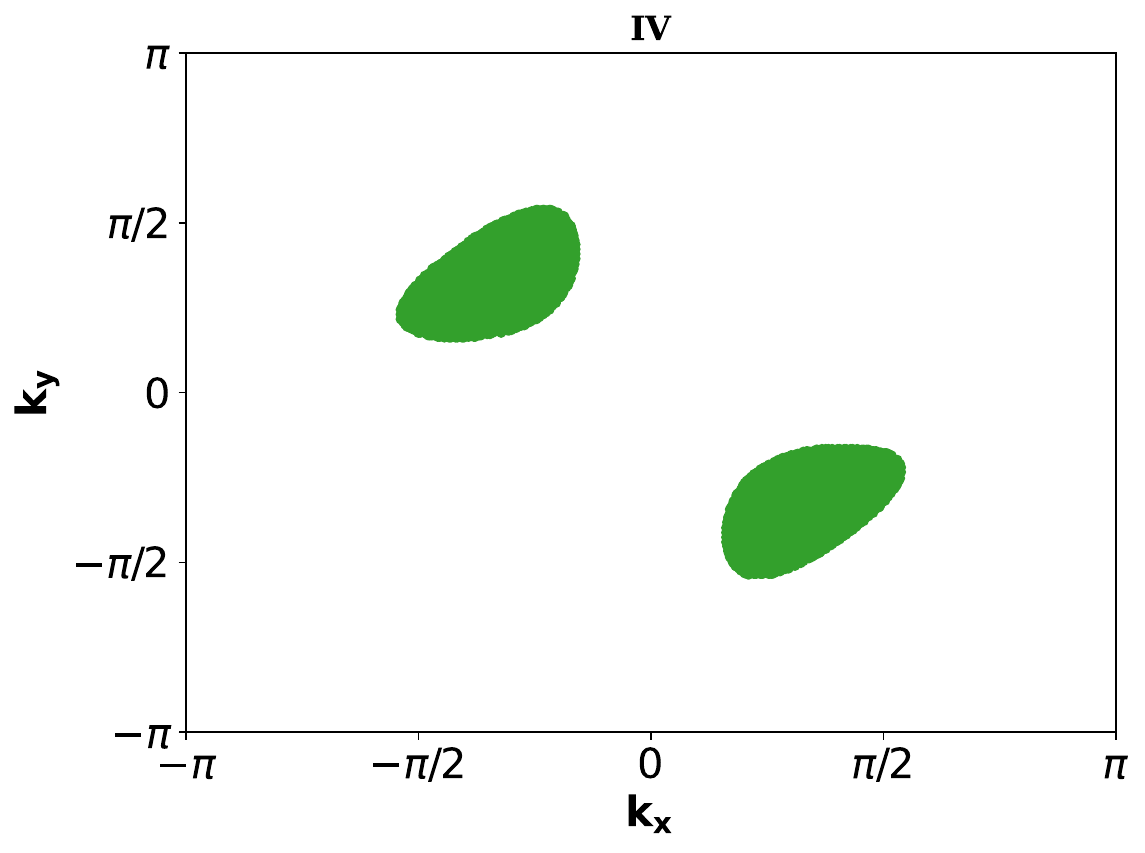}
 \includegraphics[scale=0.35]{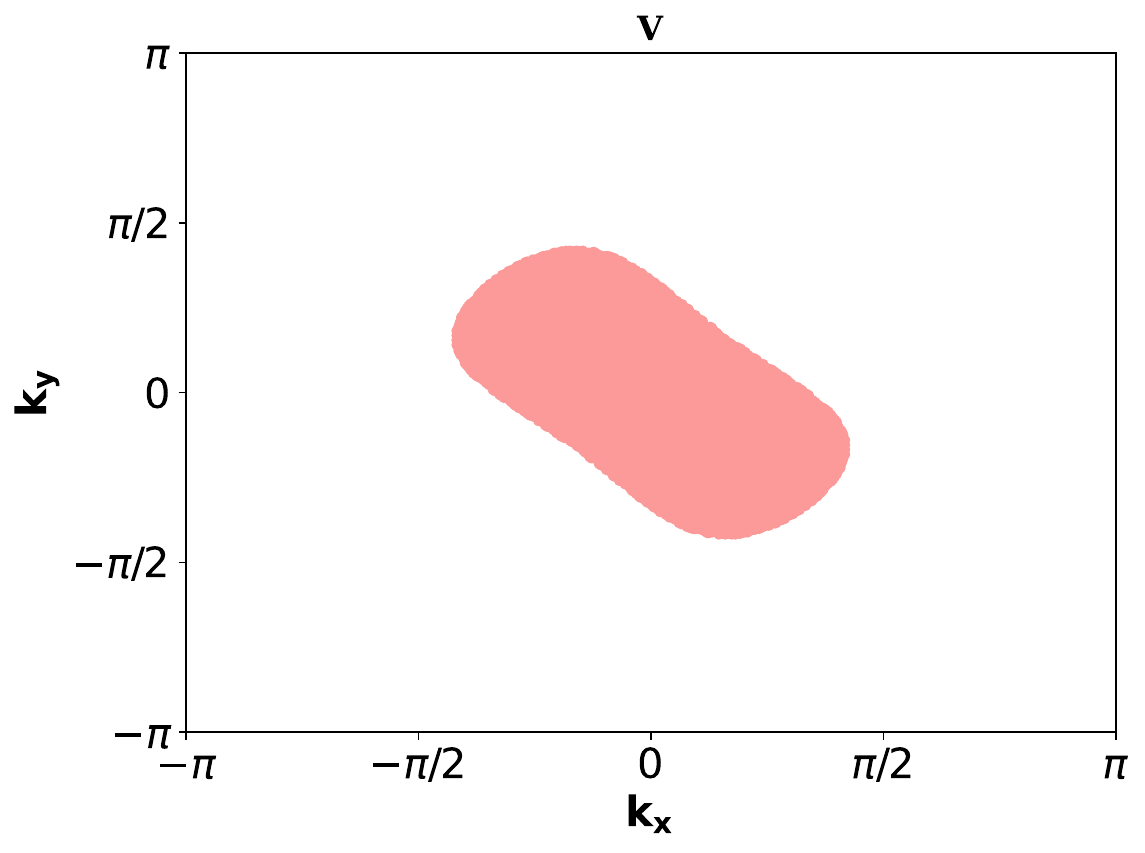}
\caption{Regions (coloured) for which the winding number is 1 on the $k_x$-$k_y$ plane with fully periodic boundaries. Note that for $J<1$, these states form a multiply connected region, and for $J>1$ they form one or more simply connected region(s). Also, these regions are the same as the regions for zero energy surface states in {Fig. 1a} under periodic boundary condition in XY plane.} 
\label{kx-ky_plane}
\end{figure*}

To calculate the winding number, we write the Hamiltonian ${\mathcal H}_{\vec{k}} $ of Eq. (\ref{def_Hk}) as
\be {\mathcal H}_{\vec{k}} = R_x \sigma_z + R_y \sigma_x \label{def_Hkw}\ee
where we introduce a  vector $\vec{R}$ with components
{\bea R_x & = & J - t\cos k_x - t \cos k_y - \cos k_z = \Delta_1 - \cos k_z
\nonumber \\ 
R_y  & = & t\sin k_x + t \sin k_y + \sin k_z = \Delta_2 + \sin k_z\eea
The last equation gives}
\be (R_x - \Delta_1)^2 + (R_y - \Delta_2)^2 =1 \label{wn1} \ee
we can see that as $k_z$ moves from $-\pi$ to $\pi$, the tip of the vector $\vec{R}$ moves over a circle with radius 1 and centre $(\Delta_1, \Delta_2)$. Therefore,  the winding number $W$ will be $1$ or $0$ according as $\Delta_1^2 + \Delta_2^2$ is smaller or larger than 1 respectively. That the condition for $W=1$ is the same as the condition Eq. (\ref{SSH_condn}) for the existence of the surface states, indicates validity of the bulk-boundary correspondence theorem \cite{SSH}. The regions of $W=1$ are shown for different phases in Fig. \ref{kx-ky_plane}. {The $W=1$ region encompasses the whole $k_x-k_y$ plane in Phase I and in Phase VI, the $W=1$ region does not exist.} \\
The dispersion relations 
are presented in the Appendix section.\\

{\bf Discussions:} We study a system of {{\em non-interacting spinless Fermions}} described by the Hamiltonian 
\bea {\mathcal H} = \sum_{\vec{k}} \left[ \left(J - t \cos k_x - t\cos k_y - \cos k_z\right) \sigma_z + \left(  t\sin k_x + t\sin k_y + \sin k_z\right) \sigma_x \right] \label{eq-Discussion}\eea 
and show that as a function of $t$ and $J$, there appear several phases having different types of topological properties of nodal lines and zero-energy surface states. The phase boundaries are also marked by different types of non-analyticities. The principal result of this work is that  
{in one of these phases (namely, Phase I, that exists for $0<J<1$, $0<t<(1-J)/2$) the energy spectrum of the Hamiltonian Eq (\ref{H_slab}) has eigenvalues which are exactly zero over the entire  Brillouin zone. The corresponding eigenstates are found to be localized on the surfaces $n_z=1$ and $n_z=L_z$. The existence of such dispersionless (`flat-band') surface states is essential for high temperature surface superconductivity}.
We mention that our Hamiltonian can be realised by suitably tuning a slab of TNLSM or by stacking layers each of which are described by {a toy model of Dirac semimetal};
and that it will be interesting to analyse the behaviour of conductivity for our Hamiltonian and to look for materials which may be represented by it.\\ \\
{\bf Acknowledgements:} PN acknowledges UGC, India for financial support (Ref. No. 191620072523) and Harish Chandra Research Institute for access to their infrastructure.
\noindent
\section*{Appendix}
\begin{figure*}
     \centering
     \begin{subfigure}[b]{0.3\textwidth}
         \centering
         \includegraphics[width=\textwidth]{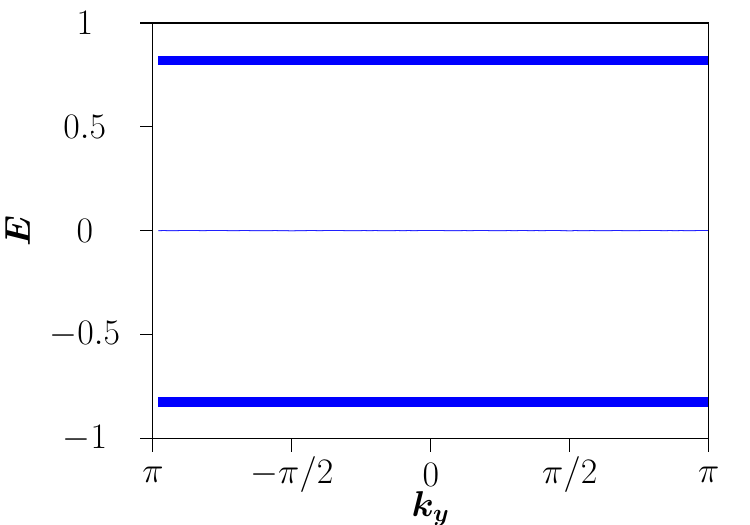}
         \caption{$k_x=0$}
         \label{}
     \end{subfigure}
     \begin{subfigure}[b]{0.3\textwidth}
         \centering
         \includegraphics[width=\textwidth]{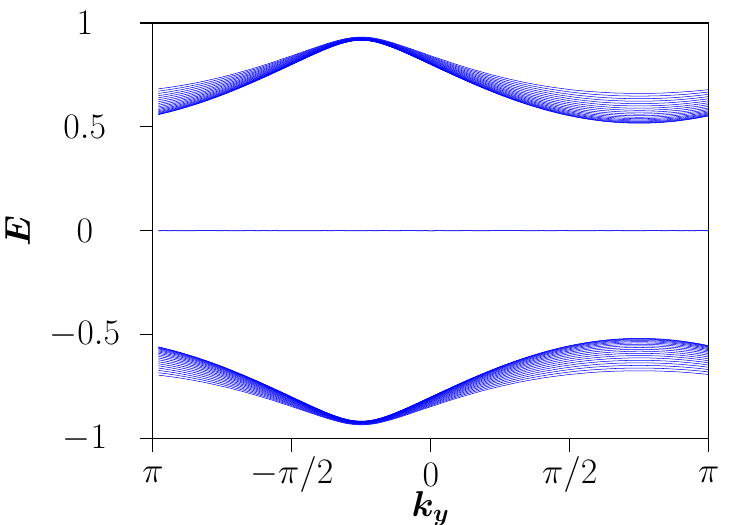}
         \caption{$k_x=\pi/2$}
         \label{}
     \end{subfigure}
     \begin{subfigure}[b]{0.3\textwidth}
         \centering
         \includegraphics[width=\textwidth]{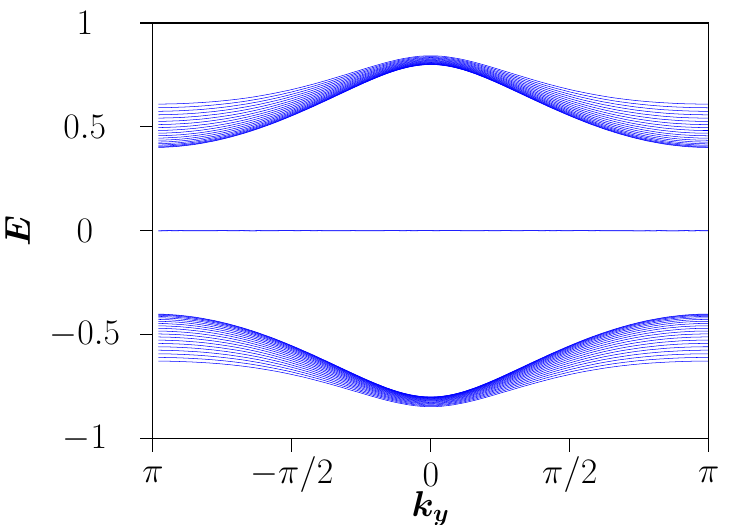}
         \caption{$k_x=\pi$}
         \label{}
     \end{subfigure}
      \begin{subfigure}[b]{0.4\textwidth}
         \centering
         \includegraphics[width=\textwidth]{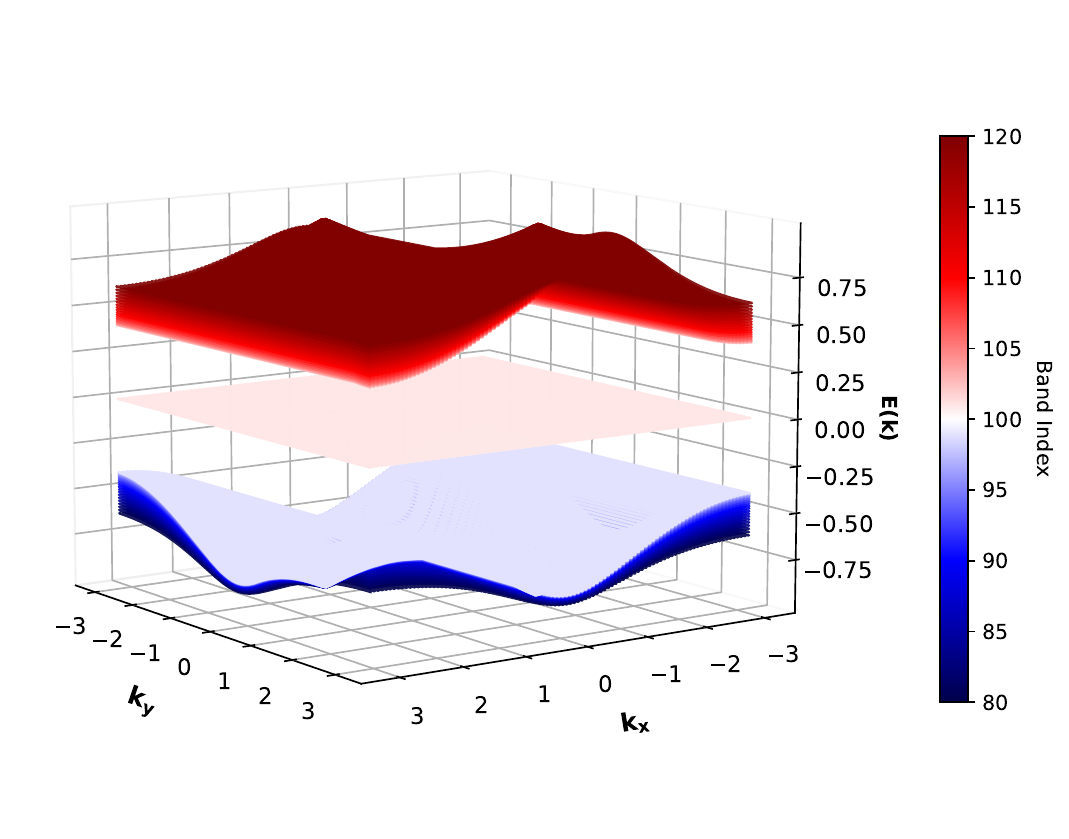}
         \caption{}
         \label{}
     \end{subfigure}
     \caption{Dispersion relation for Phase I : The dispersion relations are calculated for $J=0.2$ and $t=0.2$ in (a),(b) and (c) . We have also added the plot of the  dispersion relation over the whole $k_x-k_y$ plane in (d).}
     \label{PhaseI}
\end{figure*}

\begin{figure*}[h!]
     \centering
     \begin{subfigure}[b]{0.3\textwidth}
         \centering
         \includegraphics[width=\textwidth]{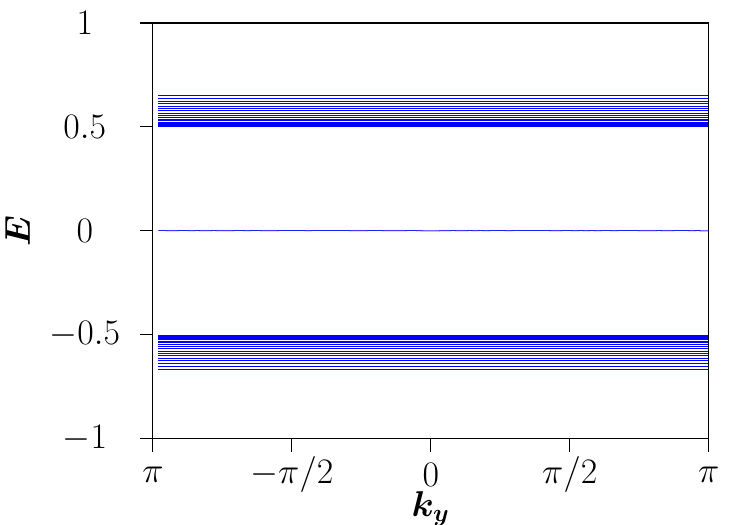}
         \caption{$k_x=0$}
         \label{}
     \end{subfigure}
     \begin{subfigure}[b]{0.3\textwidth}
         \centering
         \includegraphics[width=\textwidth]{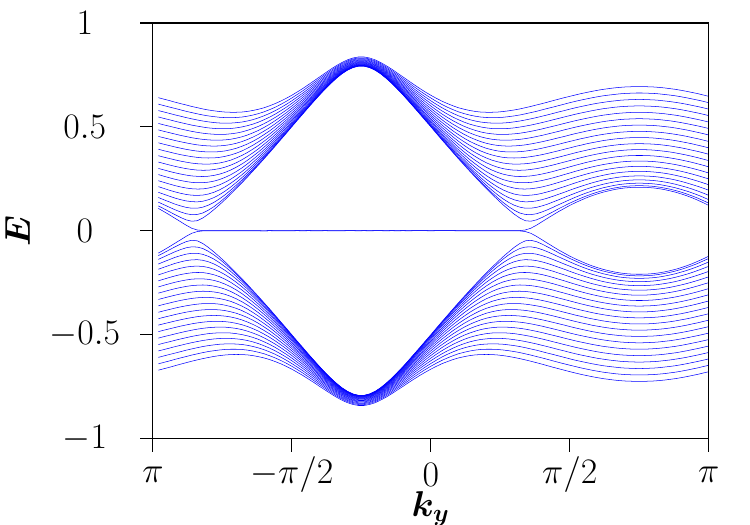}
         \caption{$k_x=\pi/2$}
         \label{}
     \end{subfigure}
     \begin{subfigure}[b]{0.3\textwidth}
         \centering
         \includegraphics[width=\textwidth]{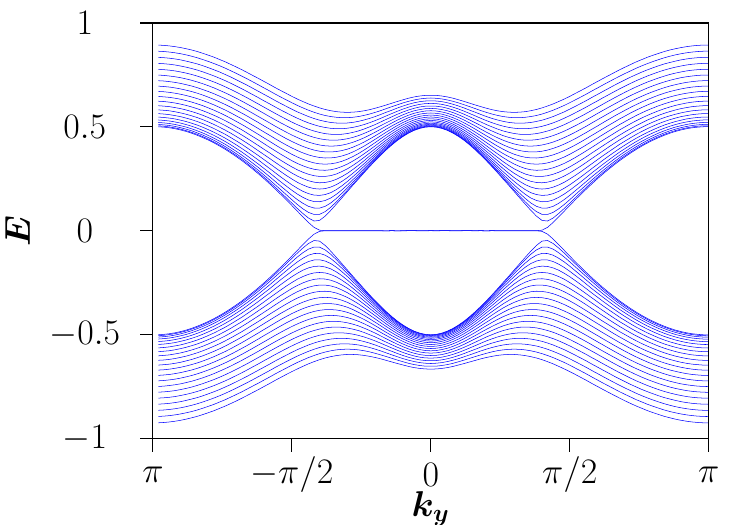}
         \caption{$k_x=\pi$}
         \label{}
     \end{subfigure}
     \caption{Dispersion relation for Phase II : The dispersion relations are calculated for $J=0.5$ and $t=0.5$}
     \label{PhaseII}
\end{figure*}

\begin{figure*}[h!]
     \centering
     \begin{subfigure}[b]{0.3\textwidth}
         \centering
         \includegraphics[width=\textwidth]{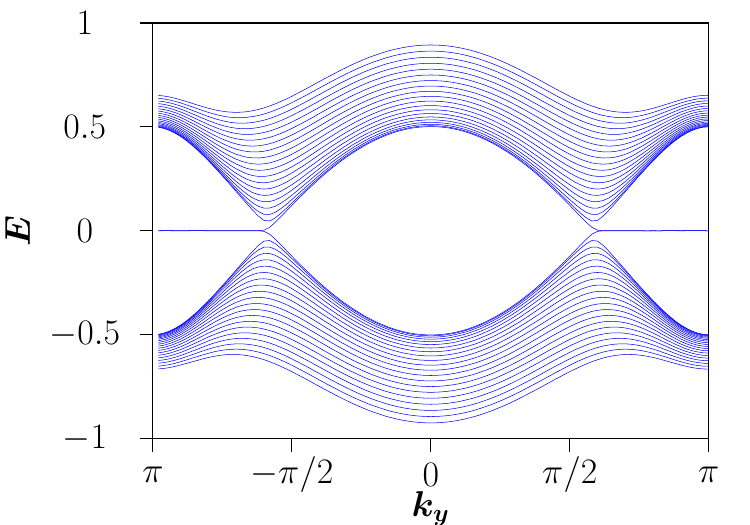}
         \caption{$k_x=0$}
         \label{}
     \end{subfigure}
     \begin{subfigure}[b]{0.3\textwidth}
         \centering
         \includegraphics[width=\textwidth]{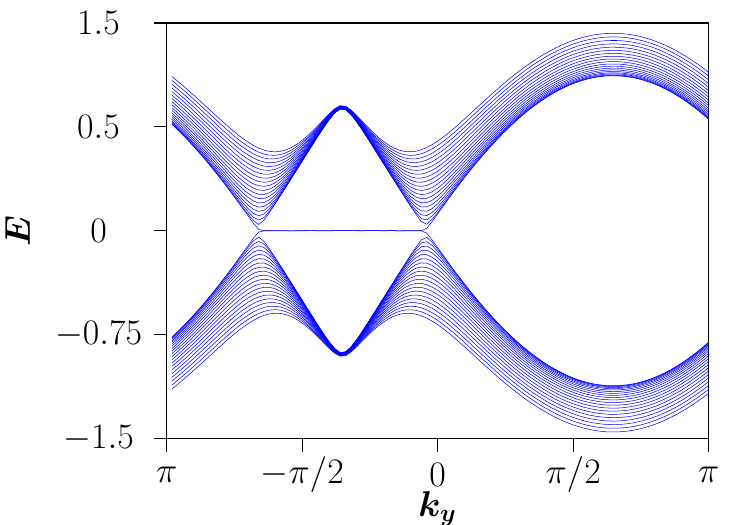}
         \caption{$k_x=\pi/2$}
         \label{}
     \end{subfigure}
     \begin{subfigure}[b]{0.3\textwidth}
         \centering
         \includegraphics[width=\textwidth]{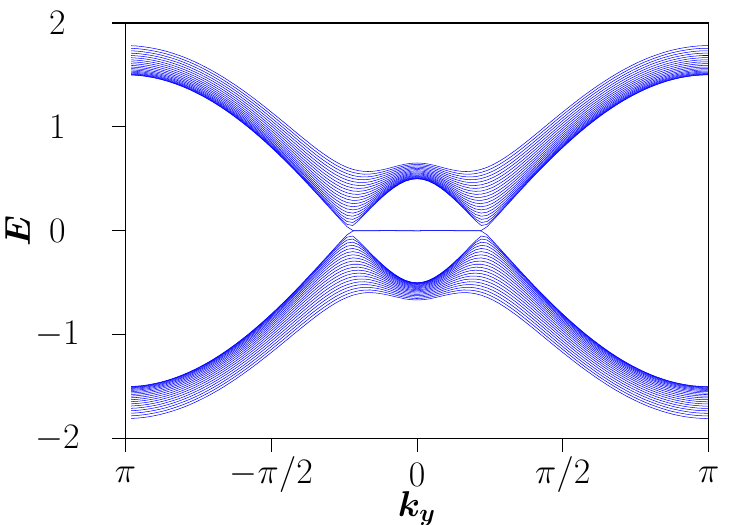}
         \caption{$k_x=\pi$}
         \label{}
     \end{subfigure}
     \caption{Dispersion relation for Phase III : The dispersion relations are calculated for $J=0.5$ and $t=1.0$}
     \label{PhaseIII}
\end{figure*}

\begin{figure*}[h!]
     \centering
     \begin{subfigure}[b]{0.3\textwidth}
         \centering
         \includegraphics[width=\textwidth]{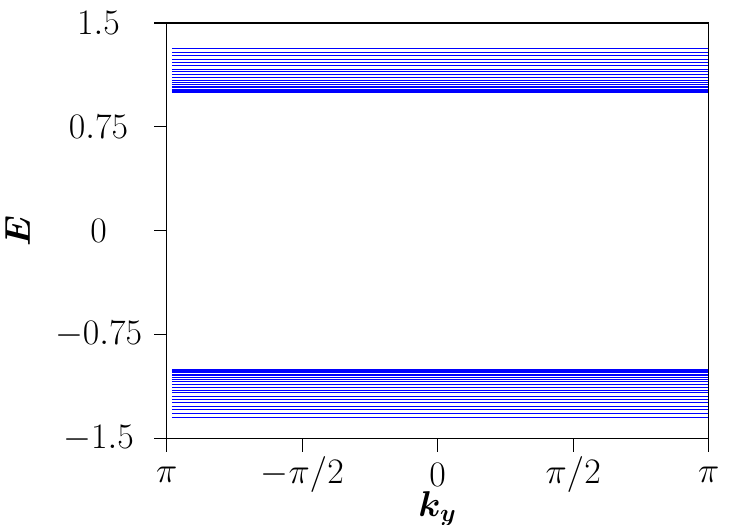}
         \caption{$k_x=0$}
         \label{}
     \end{subfigure}
     \begin{subfigure}[b]{0.3\textwidth}
         \centering
         \includegraphics[width=\textwidth]{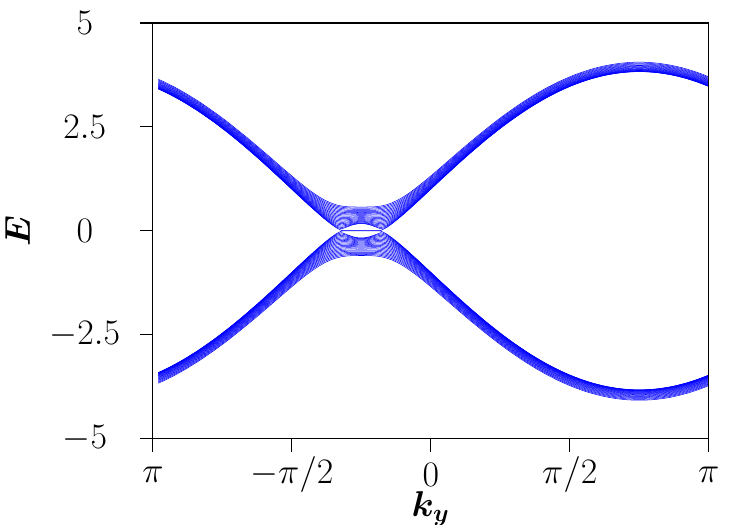}
         \caption{$k_x=\pi/2$}
         \label{}
     \end{subfigure}
     \begin{subfigure}[b]{0.3\textwidth}
         \centering
         \includegraphics[width=\textwidth]{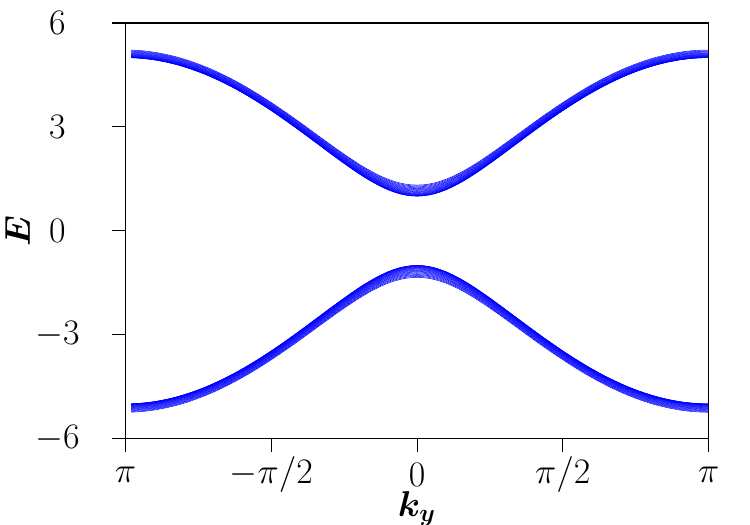}
         \caption{$k_x=\pi$}
         \label{}
     \end{subfigure}
     \caption{Dispersion relation for Phase IV : The dispersion relations are calculated for $J=2.0$ and $t=2.0$}
     \label{PhaseIV}
\end{figure*}

\begin{figure*}[h!]
     \centering
     \begin{subfigure}[b]{0.3\textwidth}
         \centering
         \includegraphics[width=\textwidth]{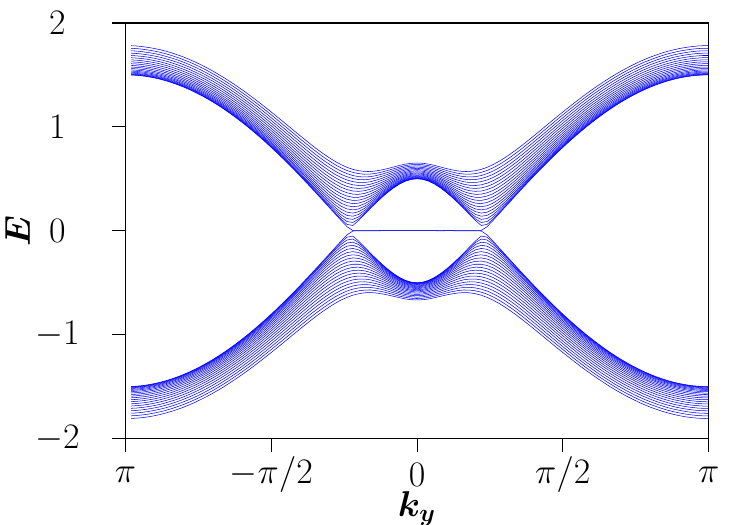}
         \caption{$k_x=0$}
         \label{}
     \end{subfigure}
     \begin{subfigure}[b]{0.3\textwidth}
         \centering
         \includegraphics[width=\textwidth]{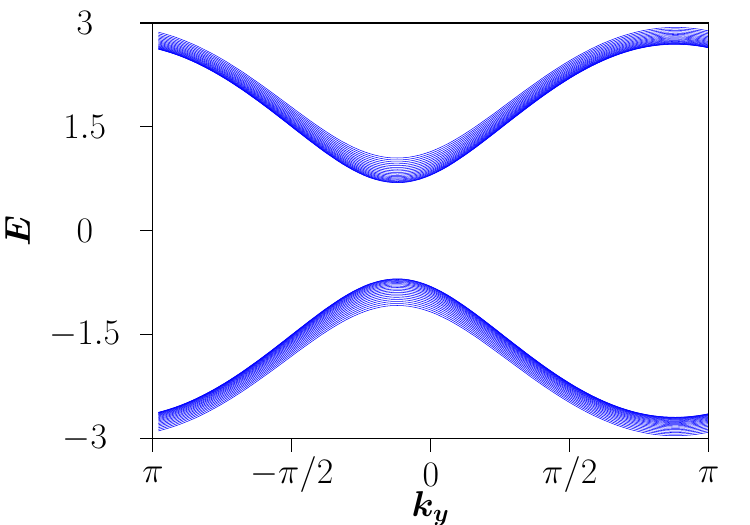}
         \caption{$k_x=\pi/2$}
         \label{}
     \end{subfigure}
     \begin{subfigure}[b]{0.3\textwidth}
         \centering
         \includegraphics[width=\textwidth]{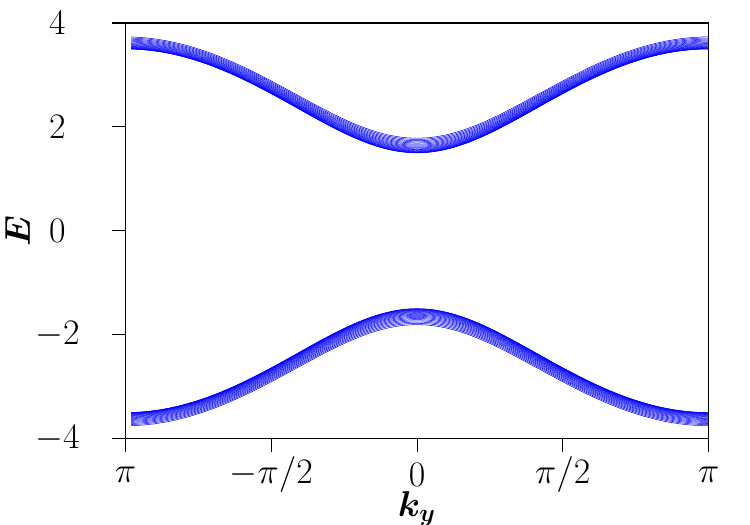}
         \caption{$k_x=\pi$}
         \label{}
     \end{subfigure}
     \caption{Dispersion relation for Phase V : The dispersion relations are calculated for $J=2.5$ and $t=1.0$}
     \label{PhaseV}
\end{figure*}

{\bf Dispersion Relation:} The dispersion relation of the effective surface Hamiltonian (Eq. \ref{H_slab}) 
is given with respect to $k_y$ for $k_x=0$, $k_x=\pi/2$ and $k_x=\pi$ in different phases in Fig (\ref{PhaseI}-\ref{PhaseV}). The calculation of the dispersion is done for $100$ layers perpendicular to the $z$-direction. The flat bands that are present in different phases correspond to the surface states, which are localized at surfaces $n_z=0$ and $n_z=100$. In the figures (\ref{PhaseI}-{\ref{PhaseV}}), we have shown only 40 bands around $E=0$. In Fig (\ref{PhaseI}), we can see that flat band surface states are present for every $k_x$ and $k_y$ in phase I. \cite{Narang2021,Rao2016} \\ 

\noindent
{\bf Surface States:} We shall derive here the condition under which the Hamiltonian (see Eq. 7 of main text)
\be H=\left( \begin {array}{cccccc}
S & \sigma^- & 0 & 0 & \cdots & 0  \\
\sigma^+ & S & \sigma^- & 0 & \cdots & 0 \\
0 & \sigma^+ & S & \sigma^- & \cdots & 0 \\
\cdots &  &  &  & & \\
0 & 0 & 0 & \cdots & \sigma^+  & S
\end{array}      \right)  \ee
admits of zero energy states at either surface $n_z=1$ or $n_z=L_z$. Here, $S = \Delta_1 \sigma_z + \Delta_2 \sigma_x $ and $\sigma^{\pm} = -(\sigma_z \pm i\sigma_x)/2$. This is an $L_z\times L_z$ matrix, each element of which is a $2 \times 2$ matrix. \\

Since this matrix has the structure of Su-Schrieffer-Heeger Hamiltonian, we shall follow Asboth \cite{SSH} and write the eigenvector of $H$ as a column matrix composed of the two-component column matrices $|x_1\rangle$,  $|x_2\rangle$, $\cdots$ $|x_{L_z}\rangle$. The energy eigenvalue equations then become
\bea
S|x_1\rangle +\sigma^-|x_2\rangle &= &E|x_1\rangle \label{eigen1}\\
\sigma^+|x_{k-1}\rangle + S|x_k\rangle + \sigma^-|x_{k+1}\rangle & = & E|x_k\rangle \nonumber\\
\;\; \text{for} \;\; 1<k<L_z \label{eigen2}  \\
\sigma^+|x_{{L_z}-1}\rangle + S|x_{L_z}\rangle & = & E|x_{L_z}\rangle \label{eigen3}
\eea
Surface states exist when these equations admit of solution with $E=0$ and either or both of $|x_1\rangle$ and $|x_{L_z}\rangle$ non-zero. We note the identities
\be \sigma^+S\sigma^-=0 \;\;\;\;\;  S^2 = \lambda^2 \underline{1}, \;\;\mbox{with }\lambda = \sqrt{\Delta_1^2 + \Delta_2^2}\ee
and put $E=0$ in Eqs (\ref{eigen1},\ref{eigen2}) to get
\be |x_k\rangle = -\frac{1}{\lambda^2}S\sigma^-|x_{k+1}\rangle \;\;\;\; 1\le k<L_z \ee
This gives,
\be \frac{1}{2}(1+\sigma_y)|x_{L_z}\rangle = \lambda^{{L_z}-1}e^{i\phi ({L_z}-1)}|x_1\rangle\ee
where $\phi$ is defined by $\Delta_1 + i\Delta_2 = \lambda e^{i\phi}$. When $\lambda<1$, one solution of this equation for large $L_z$ is $|x_{L_z}\rangle \approx 0$ even when $\langle x_1|x_1\rangle \sim 1$. Hence, Eq. (\ref{eigen3}) is satisfied, although approximately, with $E=0$, indicating the presence of zero-energy state at the edge $|x_1\rangle$. {These states will not be present when $\lambda > 1$. Thus, the condition for the existence of zero-energy states is 
\be \lambda < 1, \;\;\; {\mbox or} \;\;\; \Delta_1^2 + \Delta_2^2 < 1\ee
\noindent
{\bf Analogy with SSH model:} As mentioned above, the topological aspect of our model has resemblance with Su-Schrieffer-
Heeger model \cite{SSH}. To be more specific, we note that the Hamiltonian  for this model can be written as
\[ {\mathcal H}_k = (\Gamma + \cos k)\sigma_x + \sin k\, \sigma_y \]
where $k$ is the (one-dimensional) wave-vector and $\Gamma$ is the ratio of hopping amplitudes. The energy 
\be E(k) = \sqrt{\Gamma^2 + 1 + 2\Gamma \cos k} \label{SSH_Ek} \ee
shows a singularity at the point $\Gamma=1$, where there is a transition between topologically trivial and non-trivial phases. In our case, the  energy in Eq. (\ref{E_0(J,t)}) 
\be E(\vec{k}) \equiv \sqrt{A_{\vec{k}}^2 +B_{\vec{k}}^2 }
=\sqrt{(\Delta_1 - \cos k_z)^2 + (\Delta_2 + \sin k_z)^2}\ee
can be written as
\be E(\vec{k}) 
=\sqrt{\lambda^2 + 1 + 2\lambda \cos(k_z-\phi)} \label{Ek_last}\ee
The similarity between Eqs. (\ref{SSH_Ek}) and (\ref{Ek_last}) indicates that the topological transition and singularity at $\lambda=1$}\\

\begin{figure*}
     \centering
     \begin{subfigure}[b]{0.35\textwidth}
         \centering
         \includegraphics[width=\textwidth]{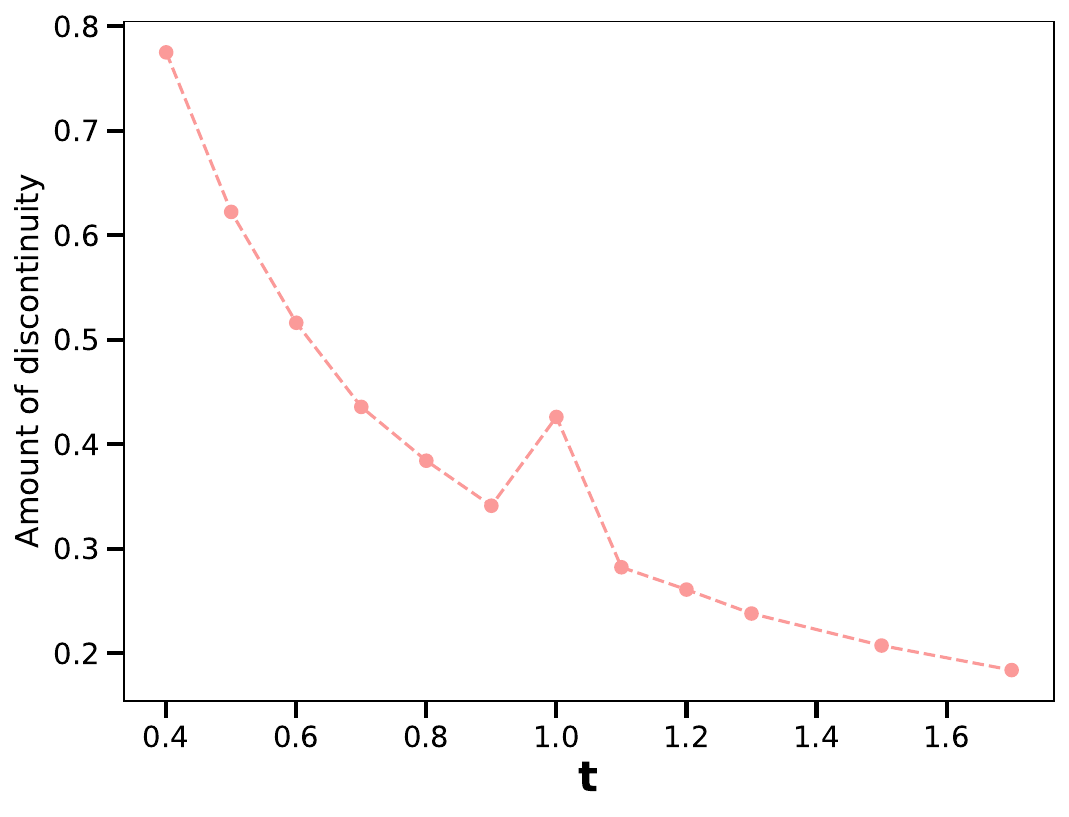}
         \caption{$J=1$ critical line}
         \label{plot_j1}
     \end{subfigure}
     \begin{subfigure}[b]{0.35\textwidth}
         \centering
         \includegraphics[width=\textwidth]{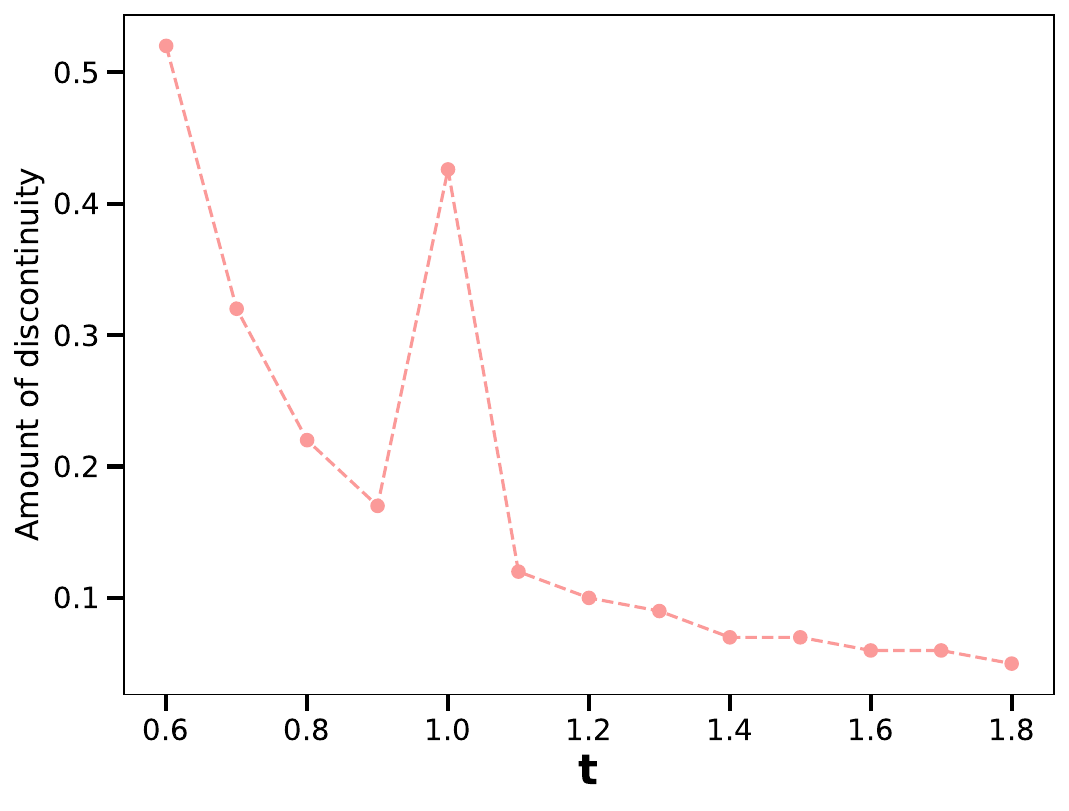}
         \caption{$J=2t-1$ critical line}
         \label{plot_j2}
     \end{subfigure}
     \caption{Amount of discontinuity vs $t$ : The behaviour of the amount of discontinuity in the third derivative of the ground state energy with the change of $t$ is shown in these plots. Note that the points lie on a smooth line apart from the point for $t=1$.}
     \label{plot_discontinuity}
\end{figure*}

\begin{figure*}
     \centering
     \begin{subfigure}[b]{0.35\textwidth}
         \centering
         \includegraphics[width=\textwidth]{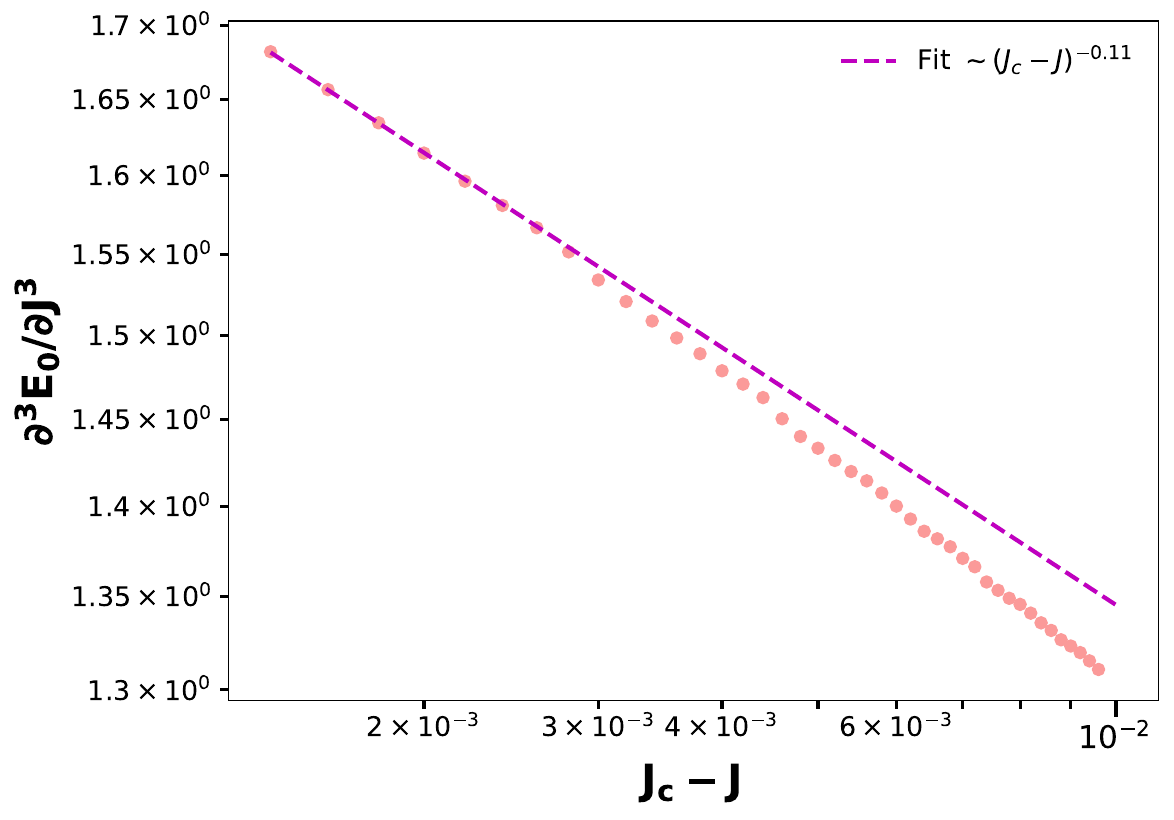}
         \caption{}
         \label{}
     \end{subfigure}
     \begin{subfigure}[b]{0.35\textwidth}
         \centering
         \includegraphics[width=\textwidth]{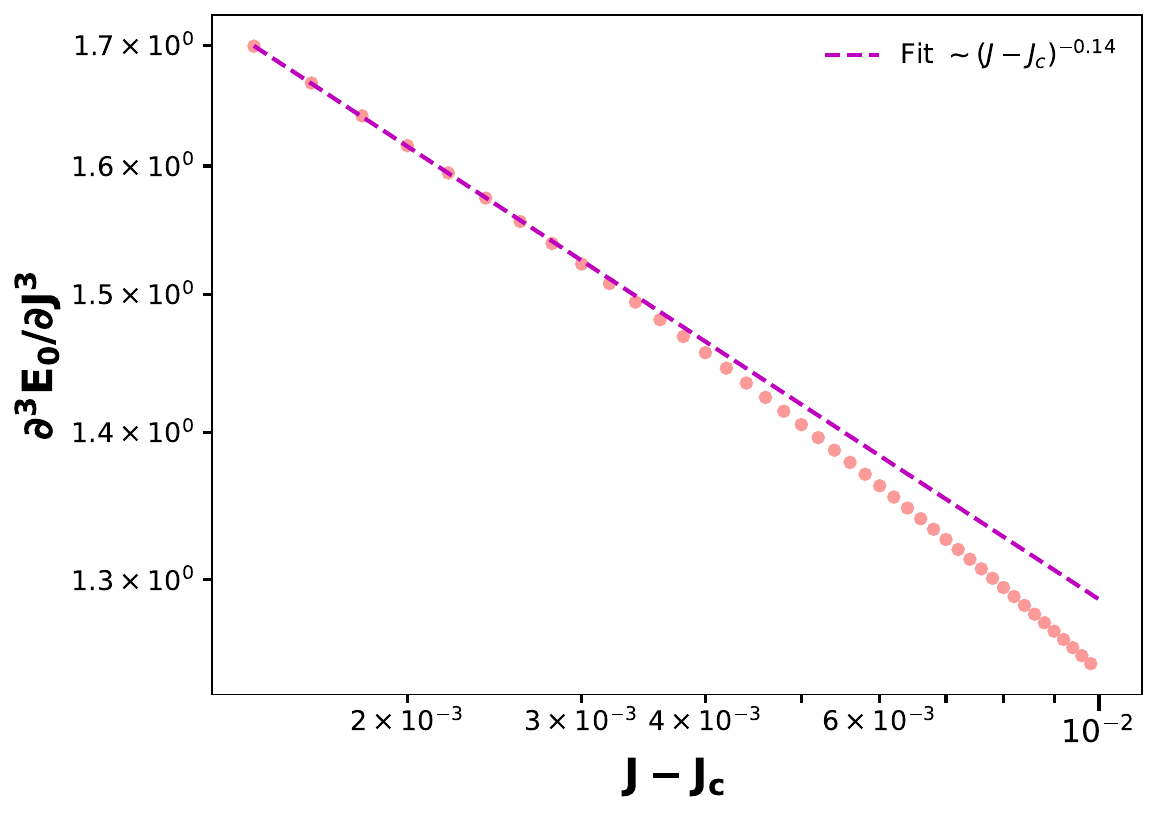}
         \caption{}
         \label{}
     \end{subfigure}
     \caption{Algebraic divergence of $\partial^3 E_0/\partial J^3$ at the critical line $J-2t=1$ separating V and VI : The dots are the numerical data and the dashed line is the fitted curve of the form $ax^{-b}$. The critical exponent is $b=0.11$ when the critical line is approached from below and $b=0.14$ when the critical line is approached from above. $t$ is fixed at $0.2$. Both $x$ and $y$ axes are in log scale.} 
     \label{cl56}
\end{figure*}

\begin{figure*}
     \centering
     \begin{subfigure}[b]{0.35\textwidth}
         \centering
         \includegraphics[width=\textwidth]{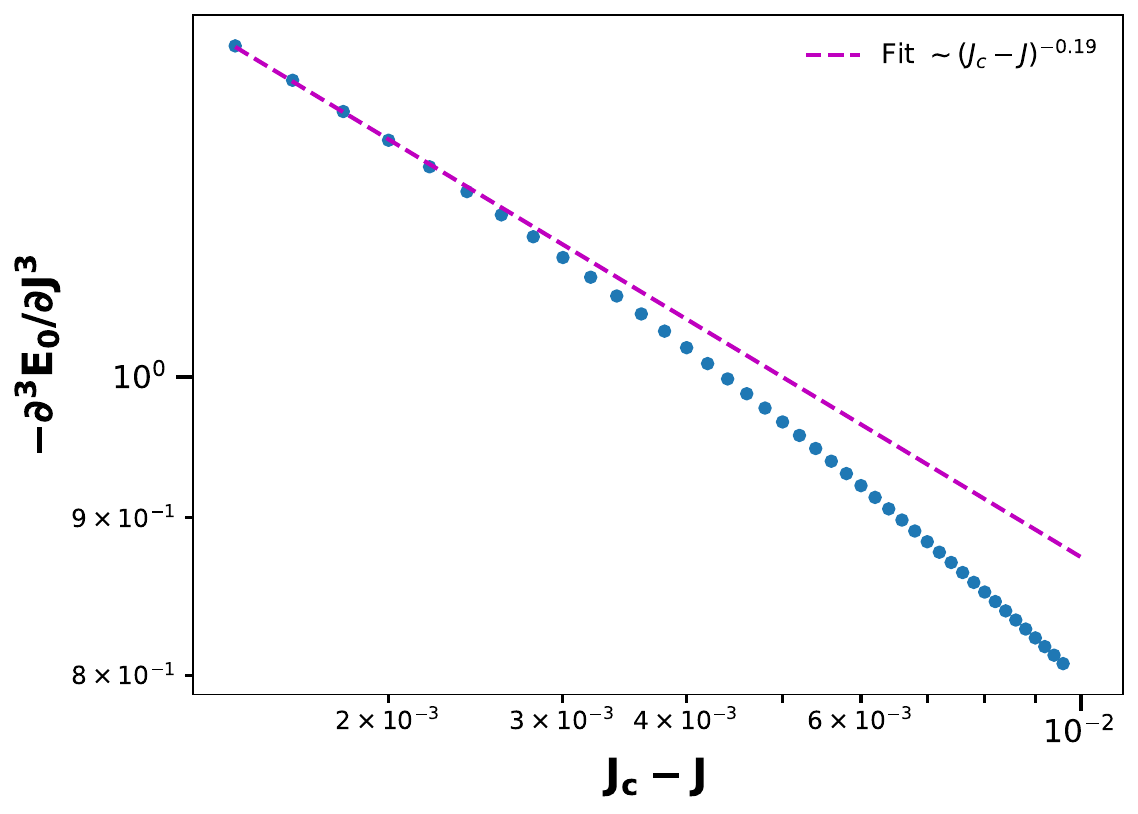}
         \caption{}
         \label{}
     \end{subfigure}
     \begin{subfigure}[b]{0.35\textwidth}
         \centering
         \includegraphics[width=\textwidth]{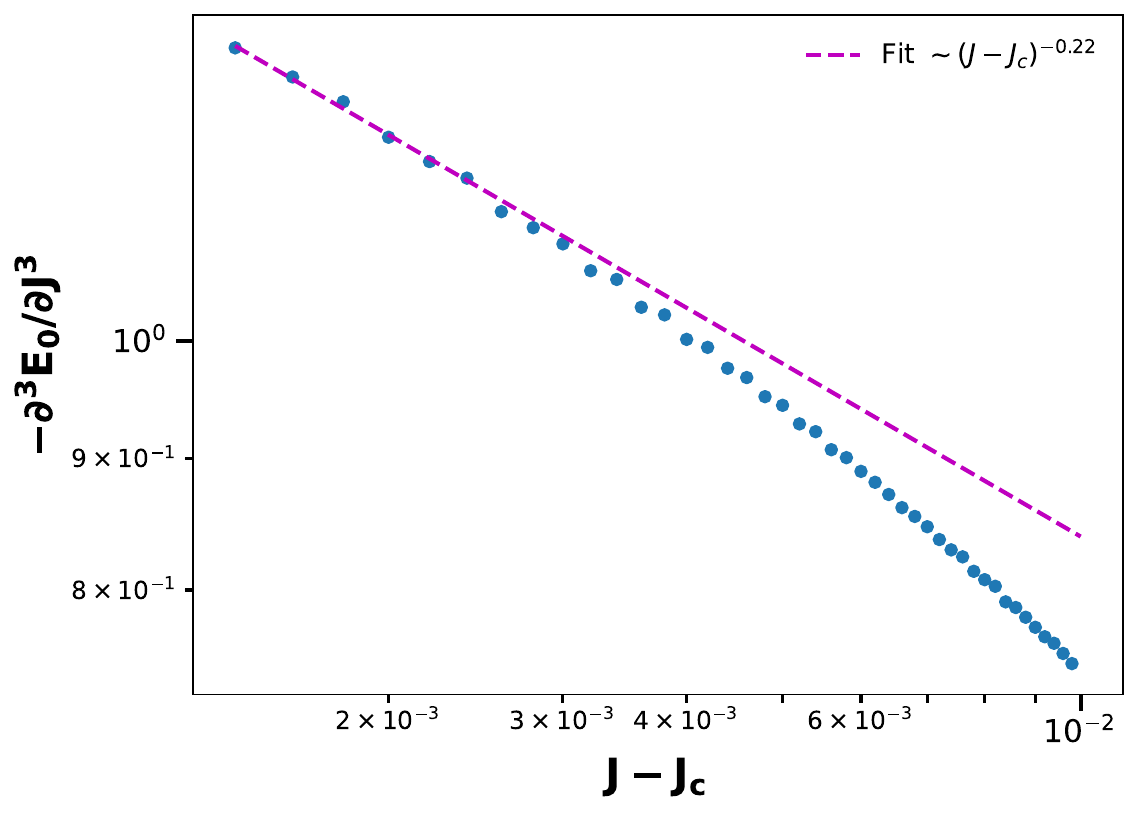}
         \caption{}
         \label{}
     \end{subfigure}
     \caption{Algebraic divergence of $\partial^3 E_0/\partial J^3$ at the critical line $J+2t=1$ separating phase I and II : The dots are the numerical data and the dashed line is the fitted curve of the form $ax^{-b}$. The critical exponent is $b=0.19$ when the critical line is approached from below and $b=0.22$ when the critical line is approached from above. $t$ is fixed at $0.3$. Both $x$ and $y$ axes are in log scale.}
     \label{cl12}
\end{figure*}

\noindent
{\bf Critical lines:} As mentioned in Fig. 1 of the main text, the non-analyticity at the phase boundaries is in the form of a discontinuity in $\partial^3 E_0/\partial t^3$ in some cases and in the form of divergence of $\partial^3 E_0/\partial t^3$ in the other cases. The amount of discontinuity varies along the boundary line. The line $J=1$ forms the boundary (II, V) and the one (III, IV). We show the variation of the amount of discontinuity along this line in Fig. (\ref{plot_j1}). Similarly, 
the line $J - 2t = -1$ forms the boundary (II, III) and the one (IV, V). The variation of the amount of discontinuity along this line is shown in Fig. (\ref{plot_j2}). On the other hand, the third derivative  $\partial^3 E_0/\partial t^3$ diverges algebraically along the boundaries (I, II) described by equation $J+2t=1$  and (V, VI) described by equation $J-2t=1$, with the exponent depending upon the direction of approach. This is shown in Figs. (\ref{cl56}, \ref{cl12}). {It needs to be mentioned that non-analyticities arise in the third derivative of the ground state energy and first and second derivatives does not have non-analyticities. So these phase transitions are neither first nor second order phase transitions \cite{Wu_2023}.} \\

\begin{figure*}[h]
     \centering
     \begin{subfigure}[b]{0.3\textwidth}
         \centering
         \includegraphics[width=\textwidth]{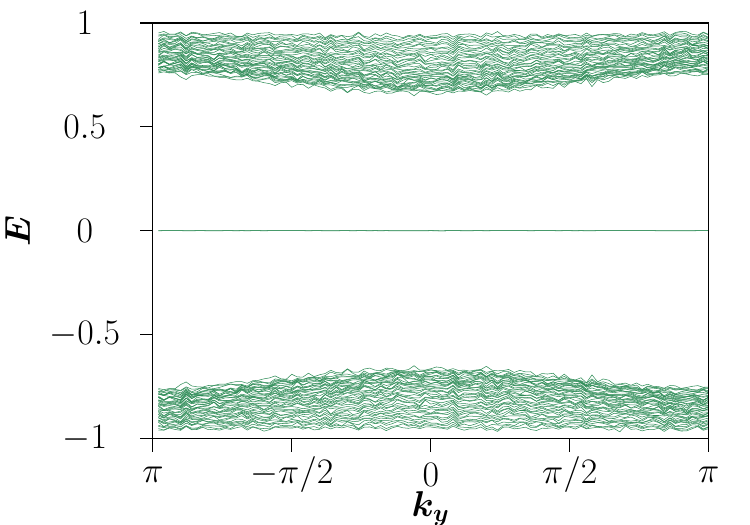}
         \caption{$k_x=0$}
         \label{}
     \end{subfigure}
     \begin{subfigure}[b]{0.3\textwidth}
         \centering
         \includegraphics[width=\textwidth]{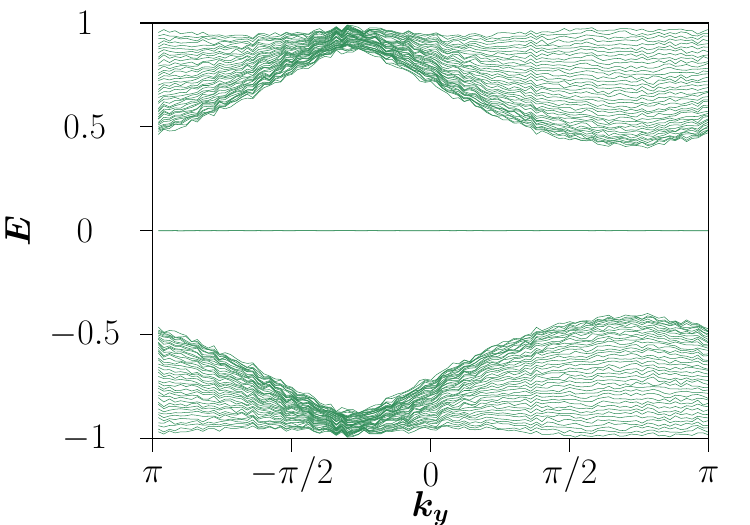}
         \caption{$k_x=\pi/2$}
         \label{}
     \end{subfigure}
     \begin{subfigure}[b]{0.3\textwidth}
         \centering
         \includegraphics[width=\textwidth]{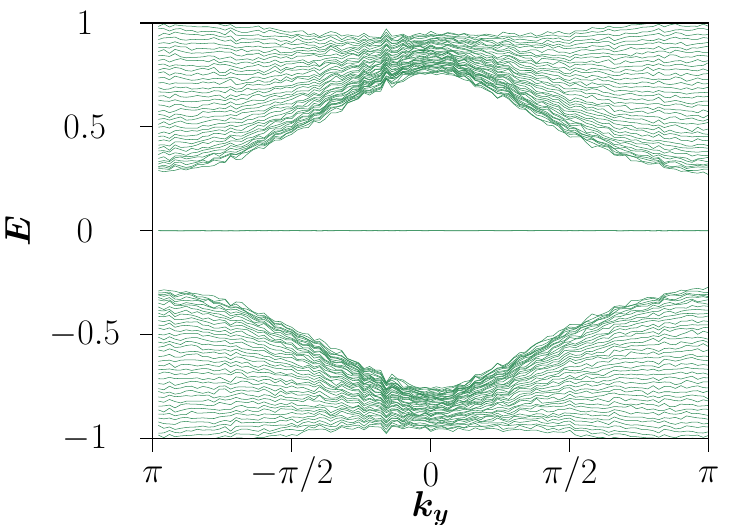}
         \caption{$k_x=\pi$}
         \label{}
     \end{subfigure}
     \caption{Dispersion relation for Phase I : The dispersion relations are calculated for $J=0.2$ and $t=0.25$. The disorder is put in the Z direction, where it fluctuates uniformly and randomly between $0.9$ and $1.1$.}
     \label{disorder_dipsersion}
\end{figure*}

{\bf Dispersion relations of Phase I with disorder:} We find that a random fluctuation of the hopping amplitude around 1.0 by 10\% does not destroy the zero energy surface states (Fig. \ref{disorder_dipsersion}).\\

\begin{figure}[h!]
     \centering
     \begin{subfigure}[b]{0.32\textwidth}
         \centering
    \includegraphics[width=\textwidth]{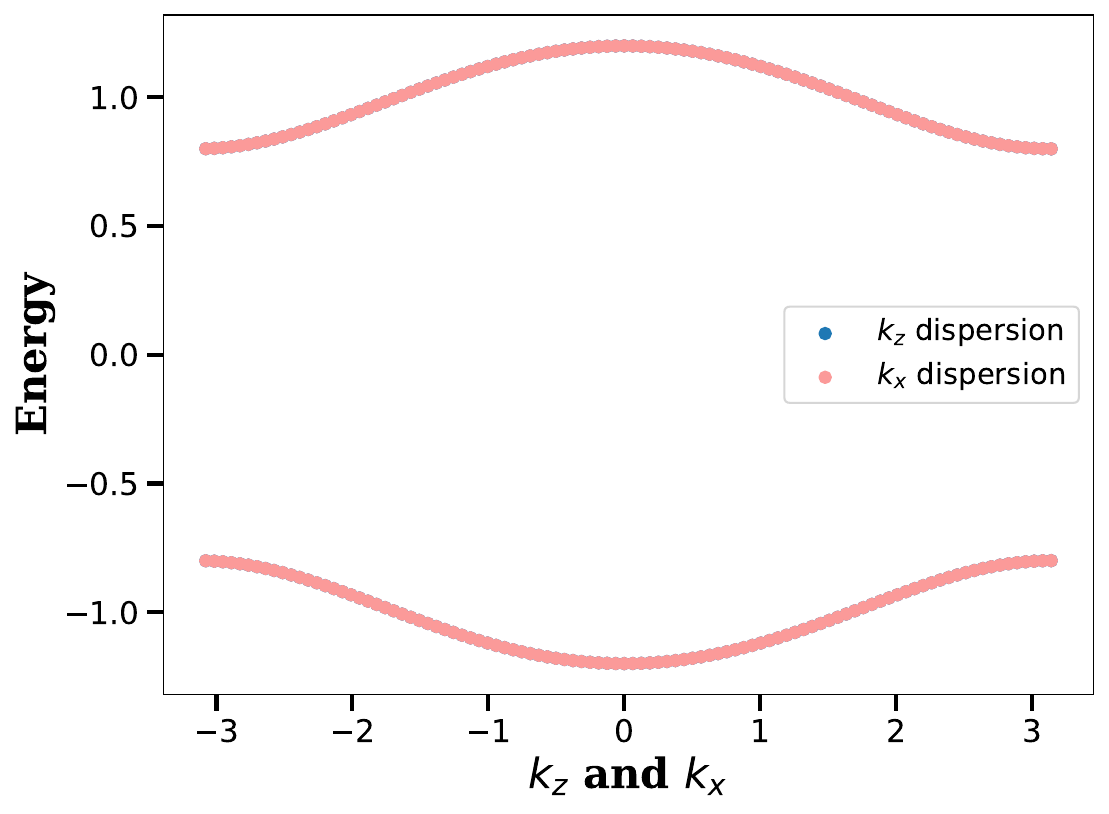}
         \caption{$k_x=k_y=0$ and $k_y=k_z=0$}
     \end{subfigure}
     \hfill
     \begin{subfigure}[b]{0.32\textwidth}
         \centering
         \includegraphics[width=\textwidth]{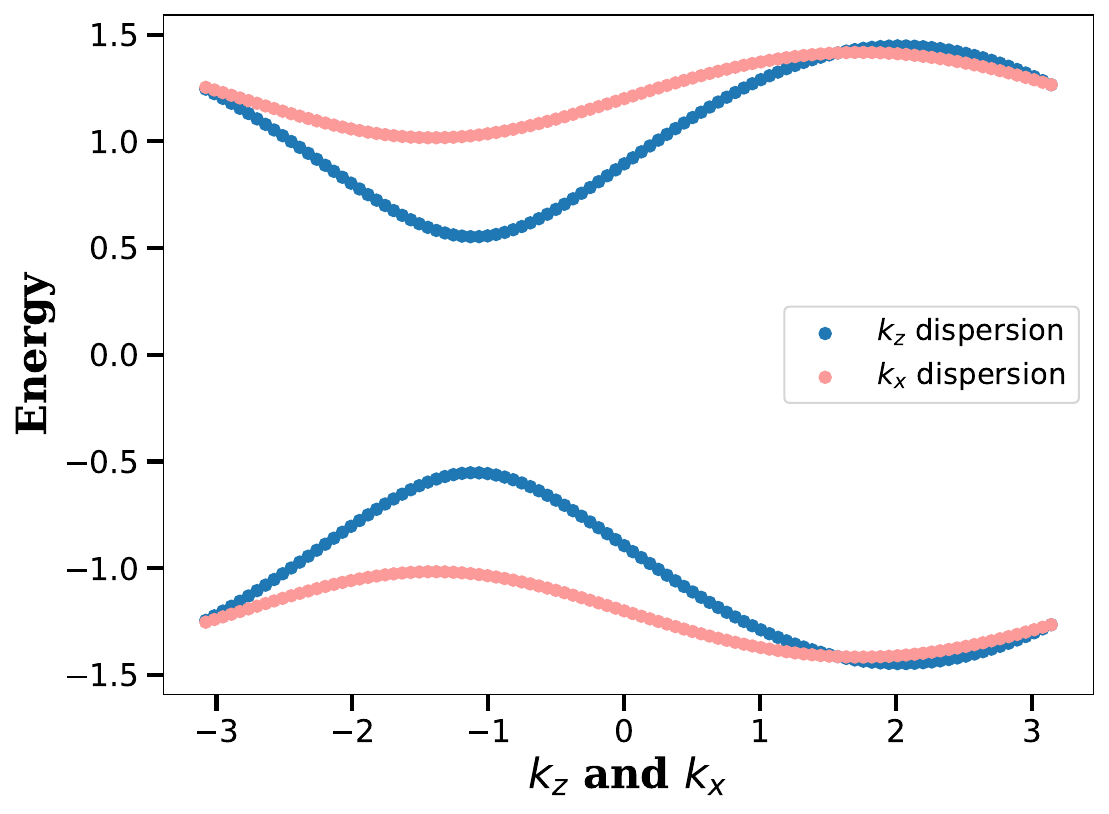}
         \caption{$k_x=k_y={\pi\over 2}$ and $k_y=k_z={\pi \over 2}$}
         \label{}
     \end{subfigure}
       \begin{subfigure}[b]{0.32\textwidth}
         \centering
         \includegraphics[width=\textwidth]{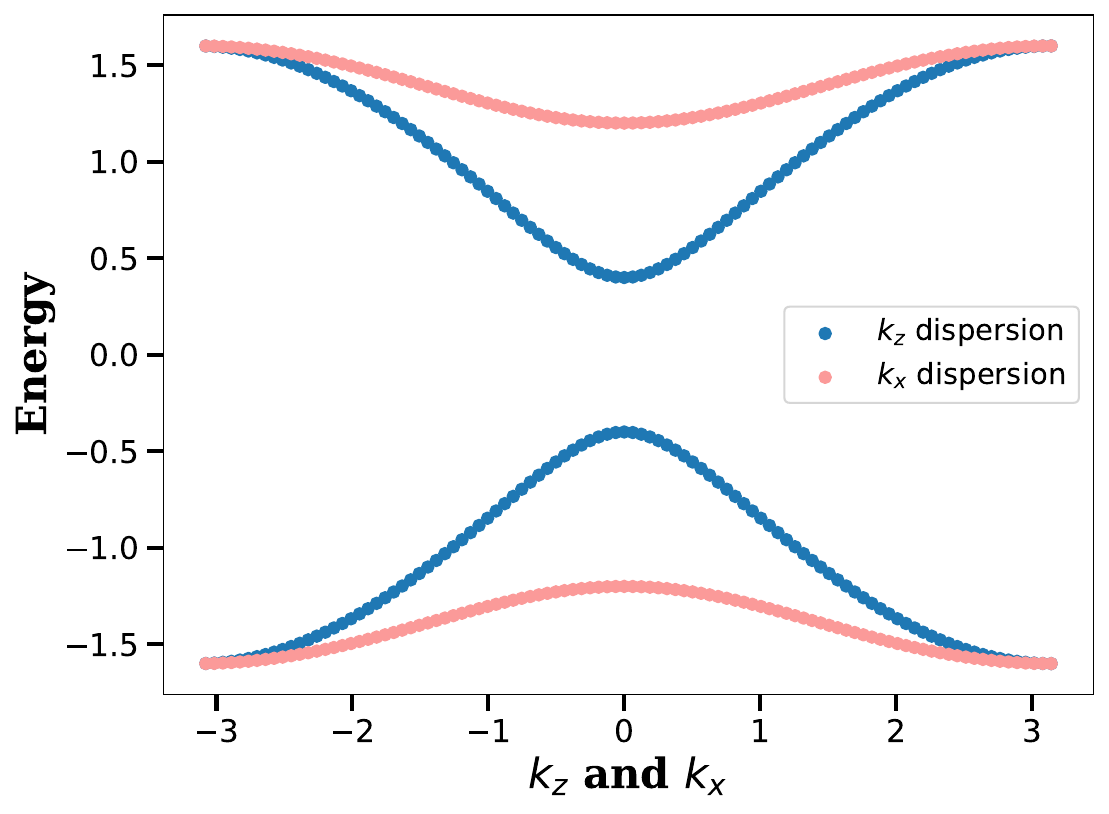}
         \caption{$k_x=k_y=\pi$ and $k_y=k_z=\pi$}
         \label{}
     \end{subfigure}
      \caption{Dispersion relation for Phase I of the Bulk Hamiltonian Eq (\ref{def_Hk})}
    \label{bulk_dipsersion}
\end{figure}

 {\bf Bulk dispersion relation of Phase I:} Here we present the dispersion relation for the bulk Hamiltonian in Phase I($J=0.2$ and $t=0.2$) in Fig. [\ref{bulk_dipsersion}]. 

\printbibliography

@article{Electride,
  title = {Electrides as a New Platform of Topological Materials},
  author = {Hirayama, Motoaki and Matsuishi, Satoru and Hosono, Hideo and Murakami, Shuichi},
  journal = {Phys. Rev. X},
  volume = {8},
  issue = {3},
  pages = {031067},
  numpages = {13},
  year = {2018},
  month = {Sep},
  publisher = {American Physical Society},
  doi = {10.1103/PhysRevX.8.031067},
  url = {https://link.aps.org/doi/10.1103/PhysRevX.8.031067}
}

@article{Duan2022,
  title = {Nodeless superconductivity in the topological nodal-line semimetal ${\mathrm{CaSb}}_{2}$},
  author = {Duan, Weiyin and Zhang, Jiawen and Kumar, Rohit and Su, Hang and Zhou, Yuwei and Nie, Zhiyong and Chen, Ye and Smidman, Michael and Cao, Chao and Song, Yu and Yuan, Huiqiu},
  journal = {Phys. Rev. B},
  volume = {106},
  issue = {21},
  pages = {214521},
  numpages = {6},
  year = {2022},
  month = {Dec},
  publisher = {American Physical Society},
  doi = {10.1103/PhysRevB.106.214521},
  url = {https://link.aps.org/doi/10.1103/PhysRevB.106.214521}
}

@article{Singh2022,
doi = {10.1088/1361-6668/ac7586},
url = {https://dx.doi.org/10.1088/1361-6668/ac7586},
year = {2022},
month = {jun},
publisher = {IOP Publishing},
volume = {35},
number = {8},
pages = {084003},
author = {M Singh and P Saha and V Nagpal and S Patnaik},
title = {Superconductivity and weak anti-localization in nodal-line semimetal SnTaS2},
journal = {Superconductor Science and Technology}
}

@article{Hirai2022,
author = {Hirai ,Daigorou and Ikenobe ,Toshiya and Yamada ,Takahiro and Yamane ,Hisanori and Hiroi ,Zenji},
title = {Unusual Resistive Transitions in the Nodal-Line Semimetallic Superconductor NaAlSi},
journal = {Journal of the Physical Society of Japan},
volume = {91},
number = {2},
pages = {024702},
year = {2022},
doi = {10.7566/JPSJ.91.024702},

URL = {https://doi.org/10.7566/JPSJ.91.024702},
eprint = {https://doi.org/10.7566/JPSJ.91.024702}
}

@article{Liu2022,
author = {Liu, Qi and Guo, Peng-Jie and Yue, Xiao-Yu and Yi, Zhe-Kai and Dong, Qing-Xin and Liang, Hui and Wu, Dan-Dan and Sun, Yan and Li, Qiu-Ju and Zhu, Wen-Liang and Xia, Tian-Long and Sun, Xue-Feng and Wang, Yi-Yan},
title = {Observation of Surface Superconductivity in a 3D Dirac Material},
journal = {Advanced Functional Materials},
volume = {32},
number = {51},
pages = {2208616},
doi = {https://doi.org/10.1002/adfm.202208616},
url = {https://onlinelibrary.wiley.com/doi/abs/10.1002/adfm.202208616},
eprint = {https://onlinelibrary.wiley.com/doi/pdf/10.1002/adfm.202208616},
year = {2022}
}

@article{Kuibarov2024,
  title={Evidence of superconducting Fermi arcs},
  author={Kuibarov, Andrii and Suvorov, Oleksandr and Vocaturo,Riccardo and Fedorov, Alexander and Lou, Rui and Merkwitz, Luise and Voroshnin, Vladimir and Facio, Jorge I and Koepernik, Klaus and Yaresko, Alexander and others},
  journal={Nature},
  volume={626},
  number={7998},
  pages={294--299},
  year={2024},
  publisher={Nature Publishing Group UK London}
}

@article{Yano2023,
  title={Evidence of unconventional superconductivity on the surface of the nodal semimetal CaAg1- x Pd x P},
  author={Yano, Rikizo and Nagasaka, Shota and Matsubara, Naoki and Saigusa, Kazushige and Tanda, Tsuyoshi and Ito, Seiichiro and Yamakage, Ai and Okamoto, Yoshihiko and Takenaka, Koshi and Kashiwaya, Satoshi},
  journal={Nature Communications},
  volume={14},
  number={1},
  pages={6817},
  year={2023},
  publisher={Nature Publishing Group UK London}
}

@book{SSH,
   title={A Short Course on Topological Insulators},
   ISBN={9783319256078},
   ISSN={1616-6361},
   url={http://dx.doi.org/10.1007/978-3-319-25607-8},
   DOI={10.1007/978-3-319-25607-8},
   journal={Lecture Notes in Physics},
   publisher={Springer International Publishing},
   author={Asbóth, János K. and Oroszlány, László and Pályi, András},
   year={2016} }

@article{Zhang2022,
  title={Promotion of superconductivity in magic-angle graphene multilayers},
  author={Zhang, Yiran and Polski, Robert and Lewandowski, Cyprian and Thomson, Alex and Peng, Yang and Choi, Youngjoon and Kim, Hyunjin and Watanabe, Kenji and Taniguchi, Takashi and Alicea, Jason and others},
  journal={Science},
  volume={377},
  number={6614},
  pages={1538--1543},
  year={2022},
  publisher={American Association for the Advancement of Science}
}

@article{Burkov2011,
  title = {Topological nodal semimetals},
  author = {Burkov, A. A. and Hook, M. D. and Balents, Leon},
  journal = {Phys. Rev. B},
  volume = {84},
  issue = {23},
  pages = {235126},
  numpages = {14},
  year = {2011},
  month = {Dec},
  publisher = {American Physical Society},
  doi = {10.1103/PhysRevB.84.235126},
  url = {https://link.aps.org/doi/10.1103/PhysRevB.84.235126}
}

@book{Shen,
  title={Topological insulators: Dirac Equation in Condensed Matter},
  author={Shen, Shun-Qing},
  publisher={Springer},
  url={https://link.springer.com/book/10.1007/978-981-10-4606-3}
}

@article{Armitage,
  title = {Weyl and Dirac semimetals in three-dimensional solids},
  author = {Armitage, N. P. and Mele, E. J. and Vishwanath, Ashvin},
  journal = {Rev. Mod. Phys.},
  volume = {90},
  issue = {1},
  pages = {015001},
  numpages = {57},
  year = {2018},
  month = {Jan},
  publisher = {American Physical Society},
  doi = {10.1103/RevModPhys.90.015001},
  url = {https://link.aps.org/doi/10.1103/RevModPhys.90.015001}
}

@article{Yang2022,
  title = {Quantum transport in topological nodal-line semimetals},
  author = {Yang, Min-Xue and Luo, Wei and Chen, Wei},
  journal = {Advances in Physics X},
  volume = {7},
  number = {1},
  pages = {2065216},
  year  = {2022},
  publisher = {Taylor & Francis},
  doi = {10.1080/23746149.2022.2065216},
  url = {https://doi.org/10.1080/23746149.2022.2065216},
}

@article{Bian,
  title={Topological nodal-line fermions in spin-orbit metal PbTaSe2},
  author={Bian, Guang and Chang, Tay-Rong and Sankar, Raman and Xu, Su-Yang and Zheng, Hao and Neupert, Titus and Chiu, Ching-Kai and Huang, Shin-Ming and Chang, Guoqing and Belopolski, Ilya and others},
  journal={Nature communications},
  volume={7},
  number={1},
  pages={10556},
  year={2016},
  publisher={Nature Publishing Group UK London},
  url={https://www.nature.com/articles/ncomms10556},
  doi={https://doi.org/10.1038/ncomms10556}
 }

@article{Paper1,
  title = {Detection of Quantum Phase Boundary at Finite Temperatures in Integrable Spin Models},
  author = {Nandi, Protyush and Bhattacharyya, Sirshendu and Dasgupta, Subinay},
  journal = {Phys. Rev. Lett.},
  volume = {128},
  issue = {24},
  pages = {247201},
  numpages = {6},
  year = {2022},
  month = {Jun},
  publisher = {American Physical Society},
  doi = {10.1103/PhysRevLett.128.247201},
  url = {https://link.aps.org/doi/10.1103/PhysRevLett.128.247201}
}

@article{Paper2,
  title = {Signature of quantum phase transition manifested in quantum fidelity at finite temperature},
  author = {Nandi, Protyush and Bhattacharyya, Sirshendu and Dasgupta, Subinay},
  journal = {Phys. Rev. B},
  volume = {109},
  issue = {6},
  pages = {064312},
  numpages = {17},
  year = {2024},
  month = {Feb},
  publisher = {American Physical Society},
  doi = {10.1103/PhysRevB.109.064312},
  url = {https://link.aps.org/doi/10.1103/PhysRevB.109.064312}
}

@article{Chiu,
  title = {Classification of topological quantum matter with symmetries},
  author = {Chiu, Ching-Kai and Teo, Jeffrey C. Y. and Schnyder, Andreas P. and Ryu, Shinsei},
  journal = {Rev. Mod. Phys.},
  volume = {88},
  issue = {3},
  pages = {035005},
  numpages = {63},
  year = {2016},
  month = {Aug},
  publisher = {American Physical Society},
  doi = {10.1103/RevModPhys.88.035005},
  url = {https://link.aps.org/doi/10.1103/RevModPhys.88.035005}
}

@article{Kopnin,
  title = {High-temperature surface superconductivity in topological flat-band systems},
  author = {Kopnin, N. B. and Heikkil\"a, T. T. and Volovik, G. E.},
  journal = {Phys. Rev. B},
  volume = {83},
  issue = {22},
  pages = {220503},
  numpages = {4},
  year = {2011},
  month = {Jun},
  publisher = {American Physical Society},
  doi = {10.1103/PhysRevB.83.220503},
  url = {https://link.aps.org/doi/10.1103/PhysRevB.83.220503}
}

@article{Weng,
  title = {Topological node-line semimetal in three-dimensional graphene networks},
  author = {Weng, Hongming and Liang, Yunye and Xu, Qiunan and Yu, Rui and Fang, Zhong and Dai, Xi and Kawazoe, Yoshiyuki},
  journal = {Phys. Rev. B},
  volume = {92},
  issue = {4},
  pages = {045108},
  numpages = {8},
  year = {2015},
  month = {Jul},
  publisher = {American Physical Society},
  doi = {10.1103/PhysRevB.92.045108},
  url = {https://link.aps.org/doi/10.1103/PhysRevB.92.045108}
}

@article{Yu,
  title = {Topological Node-Line Semimetal and Dirac Semimetal State in Antiperovskite ${\mathrm{Cu}}_{3}\mathrm{PdN}$},
  author = {Yu, Rui and Weng, Hongming and Fang, Zhong and Dai, Xi and Hu, Xiao},
  journal = {Phys. Rev. Lett.},
  volume = {115},
  issue = {3},
  pages = {036807},
  numpages = {5},
  year = {2015},
  month = {Jul},
  publisher = {American Physical Society},
  doi = {10.1103/PhysRevLett.115.036807},
  url = {https://link.aps.org/doi/10.1103/PhysRevLett.115.036807}
}

@article{Kim,
  title = {Dirac Line Nodes in Inversion-Symmetric Crystals},
  author = {Kim, Youngkuk and Wieder, Benjamin J. and Kane, C. L. and Rappe, Andrew M.},
  journal = {Phys. Rev. Lett.},
  volume = {115},
  issue = {3},
  pages = {036806},
  numpages = {5},
  year = {2015},
  month = {Jul},
  publisher = {American Physical Society},
  doi = {10.1103/PhysRevLett.115.036806},
  url = {https://link.aps.org/doi/10.1103/PhysRevLett.115.036806}
}

@article{Yan,
  title = {Collective modes in nodal line semimetals},
  author = {Yan, Zhongbo and Huang, Peng-Wei and Wang, Zhong},
  journal = {Phys. Rev. B},
  volume = {93},
  issue = {8},
  pages = {085138},
  numpages = {9},
  year = {2016},
  month = {Feb},
  publisher = {American Physical Society},
  doi = {10.1103/PhysRevB.93.085138},
  url = {https://link.aps.org/doi/10.1103/PhysRevB.93.085138}
}

@article{Rhim-Kim,
  title={Anisotropic density fluctuations, plasmons, and Friedel oscillations in nodal line semimetal},
  author={Rhim, Jun-Won and Kim, Yong Baek},
  journal={New J. Phys. },
  volume={18},
  number={4},
  pages={043010},
  year={2016},
  publisher={IOP Publishing}
}

@article{Kopnin2011supercur,
   author = {N B Kopnin},
   doi = {10.1134/S002136401113011X},
   issn = {1090-6487},
   issue = {1},
   journal = {JETP Letters},
   pages = {81-85},
   title = {Surface superconductivity in multilayered rhombohedral graphene: Supercurrent},
   volume = {94},
   url = {https://doi.org/10.1134/S002136401113011X},
   year = {2011}
}

@article{Heikkila2011flat,
   author = {T T Heikkilä and N B Kopnin and G E Volovik},
   doi = {10.1134/S0021364011150045},
   issn = {1090-6487},
   issue = {3},
   journal = {JETP Letters},
   pages = {233-239},
   title = {Flat bands in topological media},
   volume = {94},
   url = {https://doi.org/10.1134/S0021364011150045},
   year = {2011}
}

@article{Volovik2013media,
   author = {G E Volovik},
   doi = {10.1007/s10948-013-2221-5},
   issn = {1557-1947},
   issue = {9},
   journal = {Journal of Superconductivity and Novel Magnetism},
   pages = {2887-2890},
   title = {Flat Band in Topological Matter},
   volume = {26},
   url = {https://doi.org/10.1007/s10948-013-2221-5},
   year = {2013}
}

@article{Esquinazi2014,
   author = {P Esquinazi and T T Heikkilä and Y V Lysogorskiy and D A Tayurskii and G E Volovik},
   doi = {10.1134/S0021364014170056},
   issn = {1090-6487},
   issue = {5},
   journal = {JETP Letters},
   pages = {336-339},
   title = {On the superconductivity of graphite interfaces},
   volume = {100},
   url = {https://doi.org/10.1134/S0021364014170056},
   year = {2014}
}

@article{Tang2014,
   author = {Evelyn Tang and Liang Fu},
   doi = {10.1038/nphys3109},
   issn = {1745-2481},
   issue = {12},
   journal = {Nature Physics},
   pages = {964-969},
   title = {Strain-induced partially flat band, helical snake states and interface superconductivity in topological crystalline insulators},
   volume = {10},
   url = {https://doi.org/10.1038/nphys3109},
   year = {2014}
}

@article{Narang2021,
   author = {Prineha Narang and Christina A C Garcia and Claudia Felser},
   doi = {10.1038/s41563-020-00820-4},
   issn = {1476-4660},
   issue = {3},
   journal = {Nature Materials},
   pages = {293-300},
   title = {The topology of electronic band structures},
   volume = {20},
   url = {https://doi.org/10.1038/s41563-020-00820-4},
   year = {2021}
}

@article{Rao2016,
      title={Weyl semi-metals : a short review}, 
      author={Sumathi Rao},
      journal = {J. Ind. Inst. Sc.},
      volume = {96:2},
      pages = {1454},
      year = {2016},
      url = {http://journal.library.iisc.ernet.in/index.php/iisc/article/view/4611}
}

@article{Jiang2018,
   author = {Hui Jiang and Linhu Li and Jiangbin Gong and Shu Chen},
   doi = {10.1140/epjb/e2018-80717-5},
   issn = {1434-6036},
   issue = {5},
   journal = {The European Physical Journal B},
   pages = {75},
   title = {Characterization of Lifshitz transitions in topological nodal line semimetals},
   volume = {91},
   url = {https://doi.org/10.1140/epjb/e2018-80717-5},
   year = {2018}
}

@article{Wu_2023,
doi = {10.1088/1367-2630/acc820},
url = {https://dx.doi.org/10.1088/1367-2630/acc820},
year = {2023},
month = {apr},
publisher = {IOP Publishing},
volume = {25},
number = {4},
pages = {043006},
author = {Wu, H Y and Tzeng, Yu-Chin and Xie, Z Y and Ji, K and Yu, J F},
title = {Exploring quantum phase transitions by the cross derivative of the ground state energy},
journal = {New Journal of Physics}
}

@article{Fang_2016,
doi = {10.1088/1674-1056/25/11/117106},
url = {https://dx.doi.org/10.1088/1674-1056/25/11/117106},
year = {2016},
month = {nov},
publisher = {IOP Publishing},
volume = {25},
number = {11},
pages = {117106},
author = {Fang, Chen and Weng, Hongming and Dai, Xi and Fang, Zhong},
title = {Topological nodal line semimetals*},
journal = {Chinese Physics B}
}

@article{Ezawa2017,
  title = {Topological semimetals carrying arbitrary Hopf numbers: Fermi surface topologies of a Hopf link, Solomon's knot, trefoil knot, and other linked nodal varieties},
  author = {Ezawa, Motohiko},
  journal = {Phys. Rev. B},
  volume = {96},
  issue = {4},
  pages = {041202},
  numpages = {5},
  year = {2017},
  month = {Jul},
  publisher = {American Physical Society},
  doi = {10.1103/PhysRevB.96.041202},
  url = {https://link.aps.org/doi/10.1103/PhysRevB.96.041202}
}

@misc{nandi2023node,
      title={Nodal line semimetals through boundary}, 
      author={Protyush Nandi and Subinay Dasgupta},
      year={2023},
      eprint={2308.06033},
      archivePrefix={arXiv},
      primaryClass={cond-mat.stat-mech},
      url={https://arxiv.org/abs/2308.06033v1}, 
}

\end{document}